\documentclass[aps,prb,twocolumn,showpacs,superscriptaddress]{revtex4-2}

\usepackage{graphicx}
\usepackage{amsmath}
\usepackage{multirow}
\usepackage{tabularx}
\usepackage{color}
\usepackage[colorlinks,linkcolor=blue,anchorcolor=blue,citecolor=blue]{hyperref}

\begin{document}

\title{In-plane strain tuning multiferroicity in monolayer van der Waals NiI$_{2}$}	

\author{Xiao-sheng Ni}

\author{Dao-Xin Yao}
\email{yaodaox@mail.sysu.edu.cn}
\author{Kun Cao}
\email{caok7@mail.sysu.edu.cn}
\affiliation{Center for Neutron Science and Technology, Guangdong Provincial Key Laboratory of Magnetoelectric Physics and Devices, State Key Laboratory of Optoelectronic Materials and Technologies, School of Physics, Sun Yat-Sen University, Guangzhou, 510275, China}
\begin{abstract}

Multiferroic order with the engineered levels of strain in  monolayer NiI$_{2}$  is explored based on density functional theory calculations and Monte Carlo simulations. Through the investigation of strain-free monolayer NiI$_{2}$, we find that the first nearest neighbor and third nearest neighbor exchange interactions play an essential role in the formation of its magnetic phase diagrams. The competition of these interactions induces magnetic frustration, leading to the formation of proper-screw helimagnetic ground state. We further show that these conclusions drawing from the strain-free monolayer can be well generalized to the cases within our engineered range of strains. Notably, our calculations show that with 6\% tensile strain on the $a$-axis and 8\% compressive strain on the $b$-axis, the N\'{e}el temperature $T_N$ can be significantly enhanced to 101 K, about 5 times larger than that of the strain-free one. The strength of spontaneous electric polarizations can also be more than doubled under 8\% uniform compressive strain on both axis. Our work suggests that strain is a promising way to tune multiferroic orders in the monolayer NiI$_{2}$, with the potential to signicantly promote its transition temperatures and electric polarizations, therefore broaden the prospect of its applications in spintronics devices.

\end{abstract}

\maketitle

\section{Introduction}

Van der Waals (vdW) two-dimensional (2D) materials have attracted tremendous attention as they host novel properties in atomically thin systems, creating a wide area for fundamental research and potentital applications~\cite{1,2}. Many novel quantum phenomena, such as superconductivity, topological order and anomalous quantum Hall effect, which are promising in future quantum technologies, have also been found in vdW materials~\cite{2-1,2-2,2-3,2-4}. 
Among them, vdW materials with intrinsic magnetic and ferroelectric orders down to atomically thin layers, have received special attention, due to their potential to provide valuable platforms for the realization of nanoscale electronic devices~\cite{5,6}, such as spin field-effect  transistors~\cite{6-1,6-2} and ferroelectric random access memories~\cite{7-1,7-2,7-3,7,8}. Especially, vdW materials with intrinsic type-II multiferroicity, whose inversion symmetry is broken by magnetic orders, have created new opportunities to not only explore novel mechanisms of magnetoelectric coupling in 2D ~\cite{9}, but also construct multifunctional electronic devices.

Transition-metal dihalides MX$_{2}$ (where M is a transition metal and X is a haloge) is a class of vdW material with many multiferroic members, such as NiBr$_{2}$, MnI$_{2}$, CoI$_{2}$ and NiI$_{2}$~\cite{10,11}. Among them, NiI$_{2}$, a typical transition-metal dihalides with helical magnetic structures~\cite{12}, has attracted a lot of researchers to study its magnetic properties~\cite{13,14,15}. After type-II multiferroic ground state been confirmed in bulk NiI$_{2}$~\cite{16}, great efforts have been made to study its magnetic and ferroelectric properties in few layers form. Ju \textit{et al.} demonstrated that the NiI$_{2}$  has a  multiferroic state with a helical spin order even down to bilayer~\cite{17}. More recently, convincing evidence of the existence of multiferroic order in NiI$_{2}$ monolayer has been reported below a transition temperature $\sim$ 21 K~\cite{9}. Fumega \textit{et al.} then studied the freestanding NiI$_{2}$ monolayer based on \textit{ ab initio} calculations and revealed a locking between its helimagnetic and ferroelectric orders. However, the low transition temperature and weak ferroelectricity hinder its future applications in multifunctional devices. Strain engineering has been proven a powerful tool to tune properties of 2D materials, such as in CrI$_{3}$~\cite{20}, Fe$_{3}$GeTe$_{2}$~\cite{21}, and CrTe$_{2}$~\cite{22}. By varing applied strains, such as epitaxially growing monolayer NiI$_{2}$ on different substrates, we may manipulate its structure and magnetic orders to promote the multiferroic transition temperatures and electric polarizations. 

In this work, we study the multiferroic order of freestanding monolayer NiI$_{2}$ with engineered levels of strain using density functional theory (DFT) calculations and Monte Carlo(MC) simulations. Through the investigation of strain-free monolayer NiI$_{2}$, we find that the first nearest neighbor (NN) and third NN exchange interactions play an essential role in the formation of its magnetic phase diagrams. The competition of these interactions induces magnetic frustration, leading to the formation of proper-screw helimagnetic (HM) ground state. We further show that these conclusions drawing from the strain-free monolayer can be well generalized to the cases within our engineered range of strains. Notably, our calculations show that with 6\% tensile strain on the $a$-axis and 8\% compressive strain on the $b$-axis, the N\'{e}el temperature $T_N$ can be significantly enhanced to 101 K, about 5 times higher than that of the strain-free one. The strength of spontaneous electric polarizations can also be more than doubled under 8\% uniform compressive strain on both axises. Moreover, we construct a NiI$_{2}$ (monolayer)/Graphene (bilayer) heterostructure to simulate a more realistic strain engineering and obtain results in good agreement with that from our calculations on corresponding freestanding one. Our work suggests that strain is a promising way to tune multiferroic orders in the monolayer NiI$_{2}$, with the potential to signicanlty promote its transition temperatures and electric polarizations, therefore broaden the prospect of its applications in spintronics devices.

\begin{figure}[t]
	\includegraphics[scale=0.40]{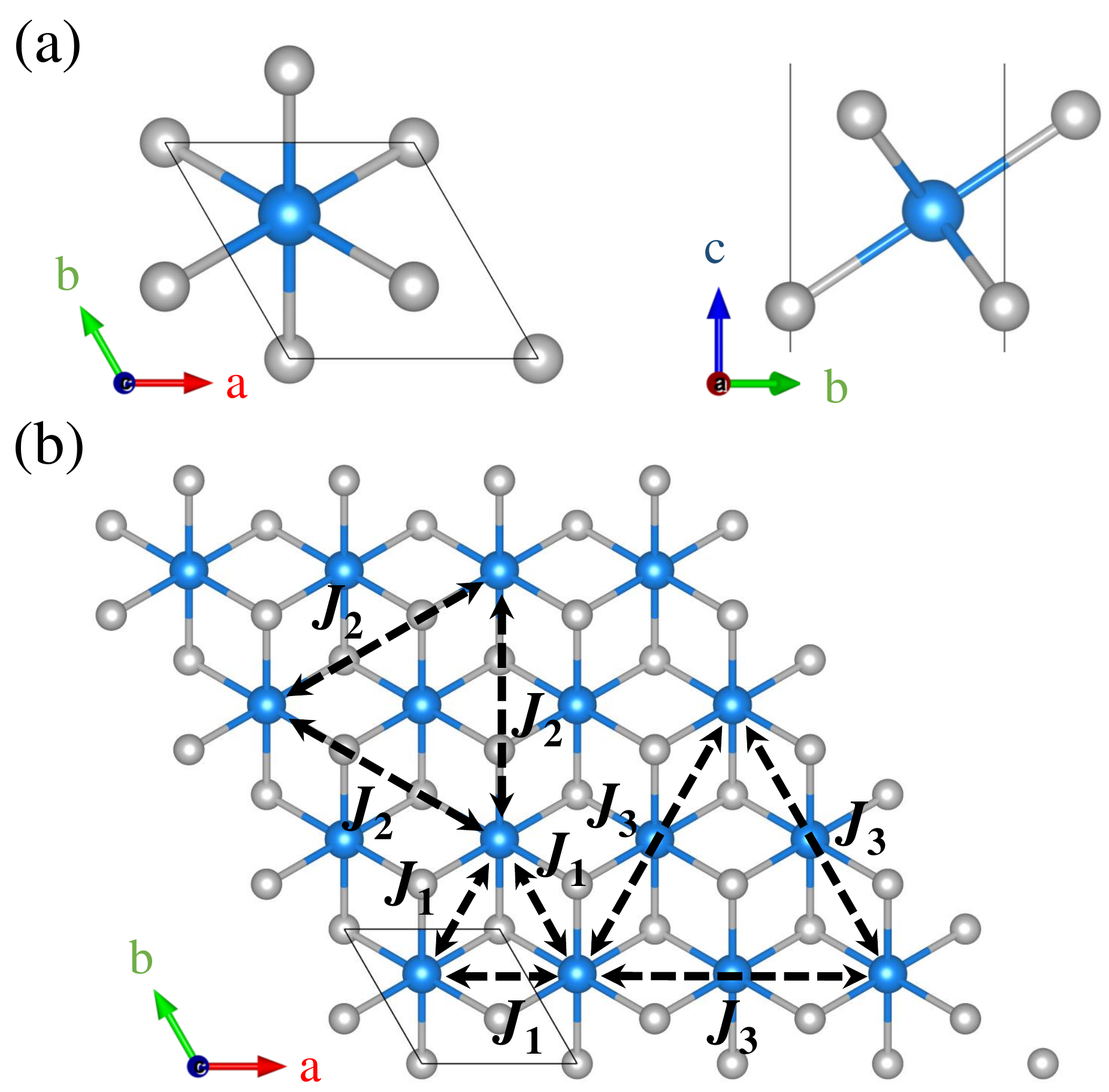}
	\caption{ (a) Side and top view of monolayer NiI$_{2}$ in a primitive cell. Blue and gray spheres represent Ni and I atoms, respectively.
	(b) The exchange paths are shown in black dotted line with arrows.}
	\label{tu1}
\end{figure}

\section{methods}

First-principles calculations are carried out with the Vienna $ab\ initio$ Simulation Package (VASP)~\cite{23,24}. We use the Perdew Burke-Ernzerhof functional with a spin-polarized generalized gradient approximation (GGA). The projector augmented-wave (PAW)~\cite{26} method with a 500 eV plane wave cutoff is employed, and a 20 $\times$ 20 $\times$ 4 $\Gamma$-centered k-point mesh allows the calculations to converge well. A vacuum layer thickness of 15 \AA\ along the out-of-plane direction is used to avoid interaction between adjacent atomic layers. The spin-polarized GGA is combined with onsite Coulomb interactions, $U$, included for Ni 3$d$ orbitals (GGA + $U$)~\cite{25}. We employ $U = 4$ eV and $J = 1$ eV, which achieves values of the magnetic moments and band gap consistent with experiments~\cite{28,29}. Upon applying strain, we relax the crystal structure until the forces acting on each atom are less than 1 meV/\AA. We determine the values of Heisenberg exchange interactions by fitting all $J$ parameters to energies calculated by DFT using 20 randomly generated collinear magnetic configurations~\cite{30,31,PhysRevB.105.214303}. In-plane strain is defined as  $\frac{a-a_0}{a_0}$ $\times$ 100\%\, where $a_0$ and $a$ denote lattice constants before and after applying strain. Moreover, we consider the strains within [ -8\%, 8\% ], which are sampled with an interval of 1\% strain along each direction. For convenience,  $x$ and $y$ are used to denote the strain level applied on the $a$-axis and $b$-axis, respectively. Electric polarizations are calculated using the Berry phase method~\cite{27}. The full atomic relaxtion for all atoms in the heterostructure takes into account the vdW force through DFT-D3 method of Grimme~\cite{grimme2010consistent}. Based on the calculated magnetic exchange interactions, we then explore the magnetic phase diagrams by a replica-exchange MC method~\cite{32}. The helical magnetic orders and corresponding magnetic propogation vectors $\boldsymbol{Q}$ are identified by performing Fourier analysis on the magnetic configurations produced from MC simulations.

\section{Results and discussions}

\subsection{Strain-free monolayer NiI$_{2}$ }

Bulk NiI$_2$ has the space group \textit{R$\overline{3}$m} (No. 166) with experimental lattice constants $a$ = $b$ = 4.46 \AA\ and c = 10.73 \AA~\cite{9,32-1}. Experiments confirm that a HM state coexists with spontaneous electric polarization in the bulk NiI$_{2}$ when tempearture drops below $T_{N}$ = 59.5 K, with a HM propogation vector $\boldsymbol{Q}$ = (0.138, 0, 1.457)~\cite{12}. The monolayer NiI$_{2}$ , which could be potentially exfoliated from a bulk NiI$_{2}$, has a reduced lattice symmetry with \textit{P$\overline{3}$m1} (No.164)~\cite{32-2}. As shown in Fig.~\ref{tu1}(a), the monolayer NiI$_{2}$ is constructed by sandwiching one layer of Ni atoms between two layers of I atoms, where the three nearest Ni atoms form an equilateral triangle, with each Ni centred at a NiI$_{6}$ octahedron. For magnetic structures, although a nearest neighbour ferromagnetic (FM) interaction is expected from a 92$^{\circ}$ Ni-I-Ni bond angle according to the Goodenough-Kanamori rule~\cite{33}, both experimental and theoretical studies show that a HM state is more stable in the bulk NiI$_{2}$~\cite{12,32-3} and even in the few-layers form~\cite{32-1,9}, suggesting that a HM ground state is also likely in the monolayer. 



\begin{figure*}[t]
	\includegraphics[scale=0.45]{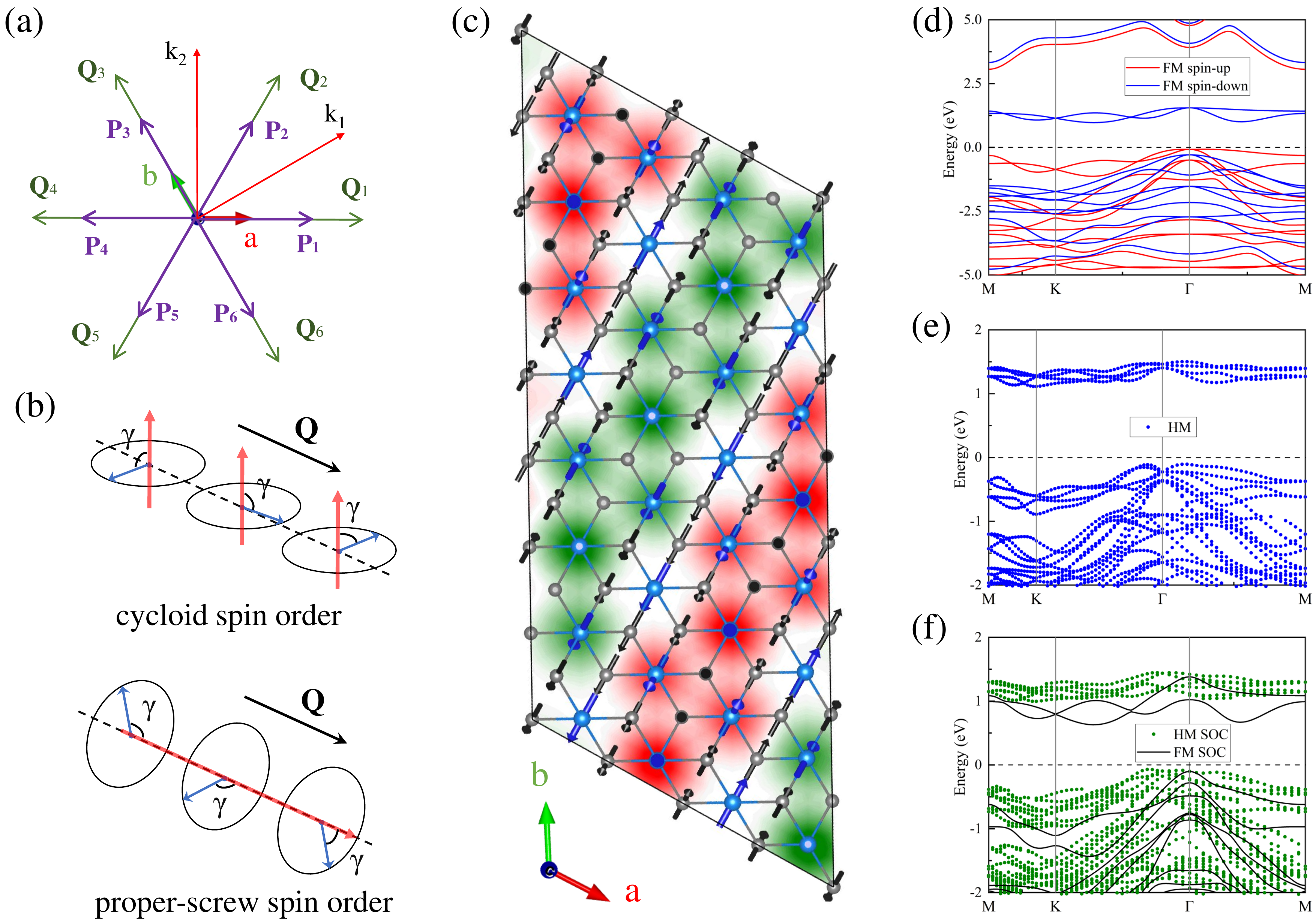}
	\caption{ (a) Schematic of the directions of $\boldsymbol{Q}$ and electric polarization $\boldsymbol{P}$ in real space. (b) The schematic of the cycloid and proper-screw magnetic order. (c) Illustration of the $\boldsymbol{Q}_{1}$ = (0.25, -0.125, 0) HM state, where the large blue arrows and small black arrows represent the spin of Ni and I atoms respectively. The colour map shows the out-of-plane spin component with red and green denoting spin-up and spin-down respectively.  (d)-(e) The band structures of monolayer NiI$_{2}$ in FM and HM states folded back into a primitive cell calculated without SOC. (f) The Band structure of FM and HM states with SOC.}
	\label{tu2}
\end{figure*}


We utilize the classic Heisenberg model to describe the magnetic interactions in freestanding monolayer NiI$_{2}$~\cite{19,34},
  \begin{equation}
    \begin{aligned}
    \textit{H} = \sum_{ij}J_{ij}\textbf{S}_i\ \cdot\ \textbf{S}_j-\sum_{i}(\textbf{D}\ \cdot\ \textbf{S}_i)^{2}
    \end{aligned}\label{eq:1}
  \end{equation}	
where $J_{ij}$ are exchange interactions between Ni spins $\boldsymbol{S}_i$ and $\boldsymbol{S}_j$. The term with $\boldsymbol{D}$ represents the magnetic anisotropy energy (MAE). We only consider exchange interactions with Ni-Ni bond lengths less than 10 \AA\ .Because of the preservation of three-fold rotation axis in the strain-free case, there are only three types of nonequivalent exchange interactions, labeled as $J_{1}$, $J_{2}$ and $J_{3}$, as shown in Fig.~\ref{tu1}(b). Our DFT calculations produce $J_{1}$, $J_{2}$ and $J_{3}$ as -4.52 meV, -0.21 meV and 3.92 meV, respectively. The $\boldsymbol{D}$ are calculated to be (0, 0, 0.39) meV, corresponding to an MAE $\sim$ 0.15 meV/Ni with an easy axis along the $c$ direction. Evidently, the MAE is too weak to affect the $T_N$, as well as the magnetic vectors of the HM states, therefore for simplicity, we first analyse the ground state magnetic structures without considering MAE. 

In the absence of MAE, a spin rotation axis in HM state can be generally assumed to be along the $c$ direction and then each spin value in the monolayer NiI$_{2}$ can be determined by
\begin{equation}
\begin{aligned}
\textbf{S$_{n,0}$} = &S_0\cdot(cos(\textbf{Q} \cdot\ (\textbf{R}_n\ + \textbf{R}_0\ ))sin(\gamma\ ))\ \textbf{a}\ \\
&+ S_0\cdot(sin(\textbf{Q} \cdot\ (\textbf{R}_n\ + \textbf{R}_0\ ))sin(\gamma\ ))\ \textbf{b}\  \\
&+ cos(\gamma\ )\ \textbf{c}\
\end{aligned}\label{eq:2}
\end{equation}
where $S_{0}$ is the magnitude of each spin, $\boldsymbol{R}_{n}$ is the lattice vector pointing from the unit cell 0 to unit cell n,  $\boldsymbol{R}_{0}$ is the position of the spin at origin, $\gamma$ is the cone angle between the rotation axis and the magnetic moment of each atoms.

Denoting $\boldsymbol{Q}$ =($q_{a}$, $q_{b}$, 0), the magnetic Hamiltonian in a primitive cell can then be explicitly written as
  \begin{equation}
\begin{aligned}
\textit{H$_{iso}$} &= J_1 S_0^2 sin^2(\gamma\ ) (cos(q_a)+cos(q_b)+cos(q_a+q_b))\ \\
&+J_2 S_0^2 sin^2(\gamma\ ) (cos(2q_a+q_b)+cos(q_a+2q_b)\\
& \ \ \  \ \ \ \ \ \  \ \ \ \ \ \  \ \ \ \ \ \ \ +cos(q_a-q_b))\ \\
&+ J_3 S_0^2 sin^2(\gamma\ ) (cos(2q_a)+cos(2q_b)\\
& \ \ \  \ \ \ \ \ \  \ \ \ \ \ \  \ \ \ \ \ \ \ +cos(2q_a+2q_b))\ \\
&+(J_1+J_2+J_3)S_0^2 cos^2(\gamma\ )
\end{aligned}\label{eq:3}
\end{equation}
By performing derivation of Eq. (\ref{eq:3}) with respect to $q_{a}$ and $q_{b}$,  we can find the corresponding magnetic ground state.
  \begin{equation}
\begin{aligned}
\frac{\partial\ \textit{H$_{iso}$}}{\partial\ q_a} =0\ \& \ \frac{\partial\ \textit{H$_{iso}$}}{\partial\ q_b} =0
\end{aligned}\label{eq:4}
\end{equation}
By solving Eq. (\ref{eq:4}), we get the relation
  \begin{equation}
\begin{aligned}
 q_a , q_b \sim arctan((\frac{J_1 +3J_2 +4J_3 }{J_2-J_1+4J_3})^{1/2})
\end{aligned}\label{eq:5}
\end{equation}
From Eq. (\ref{eq:5}), we can see that the solutions of $q_{a}$ and $q_{b}$ only exist when
  \begin{equation}
\begin{aligned}
\frac{J_1 +3J_2 +4J_3 }{J_2-J_1+4J_3} > 0
\end{aligned}\label{eq:6}
\end{equation}

\begin{figure}[t]
	\includegraphics[scale=0.50]{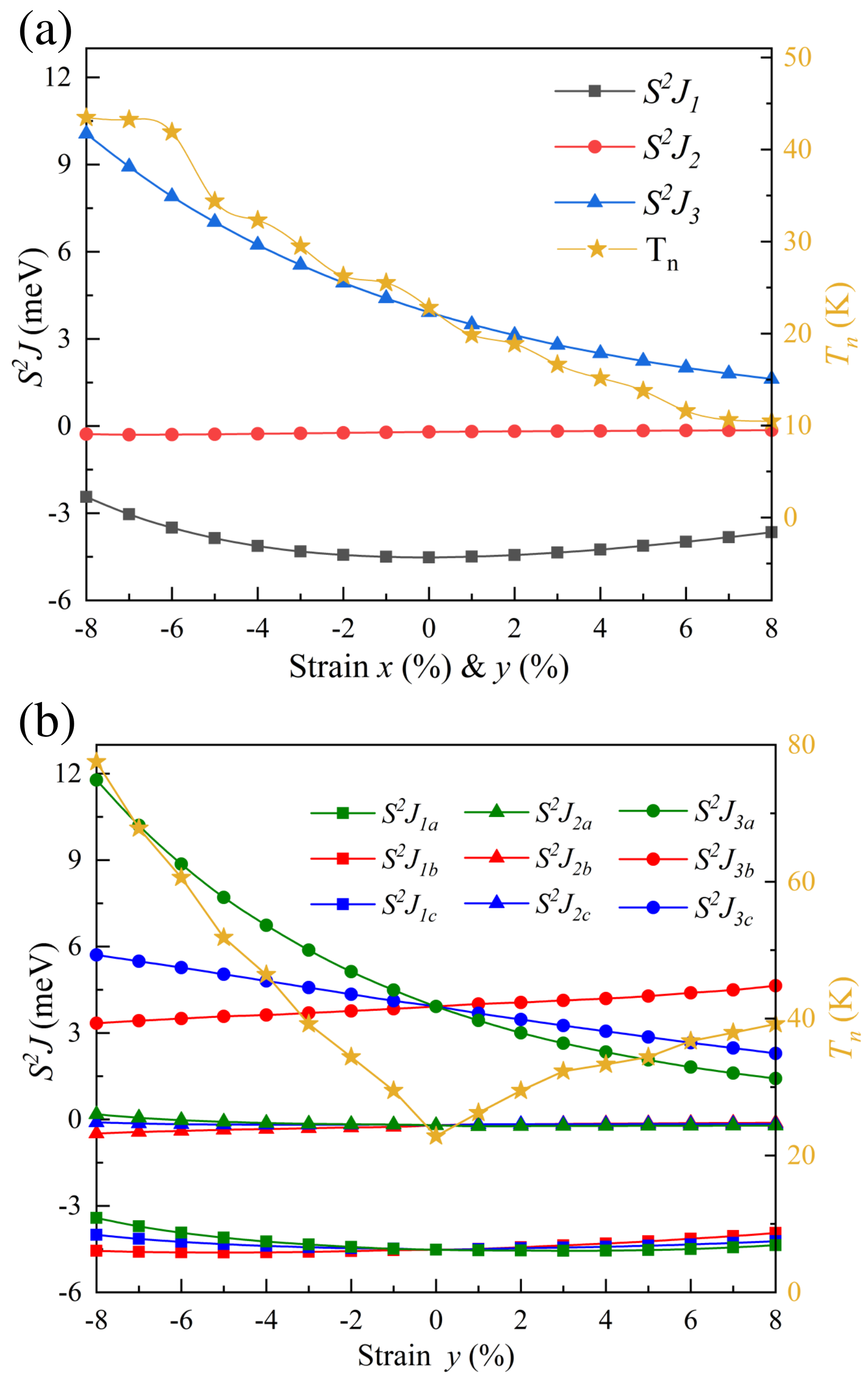}
	\caption{ (a) The magnetic exchange interactions $J_{1}$, $J_{2}$ , $J_{3}$, and the N\'{e}el temperature $T_N$ with uniform strains. (b) The $J_{1a-1c}$,$J_{2a-2c}$ , $J_{3a-3c}$ and $T_N$ as functions of $y$, with $x=0 \%$.}
	\label{tu3}
\end{figure}

\begin{table*}[t!]
	\caption{ \label{tab1-1} Values (in meV) of magnetic exchange couplings and $T_{N}$ by DFT calculation and MC simulations in three typical strains. }
	\begin{ruledtabular}
		\begin{tabular}{cccccccccccc}
			& $x$($a$-axis) \& $y$($b$-axis) strain & $ J_{1a}$ & $ J_{1b}$ & $J_{1c}$ & $ J_{2a}$  & $ J_{2b}$ & $J_{2c}$ & $ J_{3a}$ & $ J_{3b}$ & $ J_3c$ & $T_{N}$(K) \\
			\hline  \\[-1.0ex]
			& 7\% \& 8\% & -3.74 & -3.73 & -3.74 & -0.14 & -0.14 & -0.16 & 1.83 & 1.70 & 1.59 & 13.77  \\
			& 7\% \& -3\% &-4.02 & -4.39 & -4.60 & -0.06 & -0.16 & -0.26 & 6.74 & 2.76 & 1.53  & 51.85\\
			& 6\% \& -8\% &-3.42 & -4.25 & -4.68 & 0.34 & -0.05 & -0.44 & 13.27 & 3.47 & 1.52 &  101.17 \\
		\end{tabular}
	\end{ruledtabular}
\end{table*}

\begin{figure}[t]
	\includegraphics[scale=0.47]{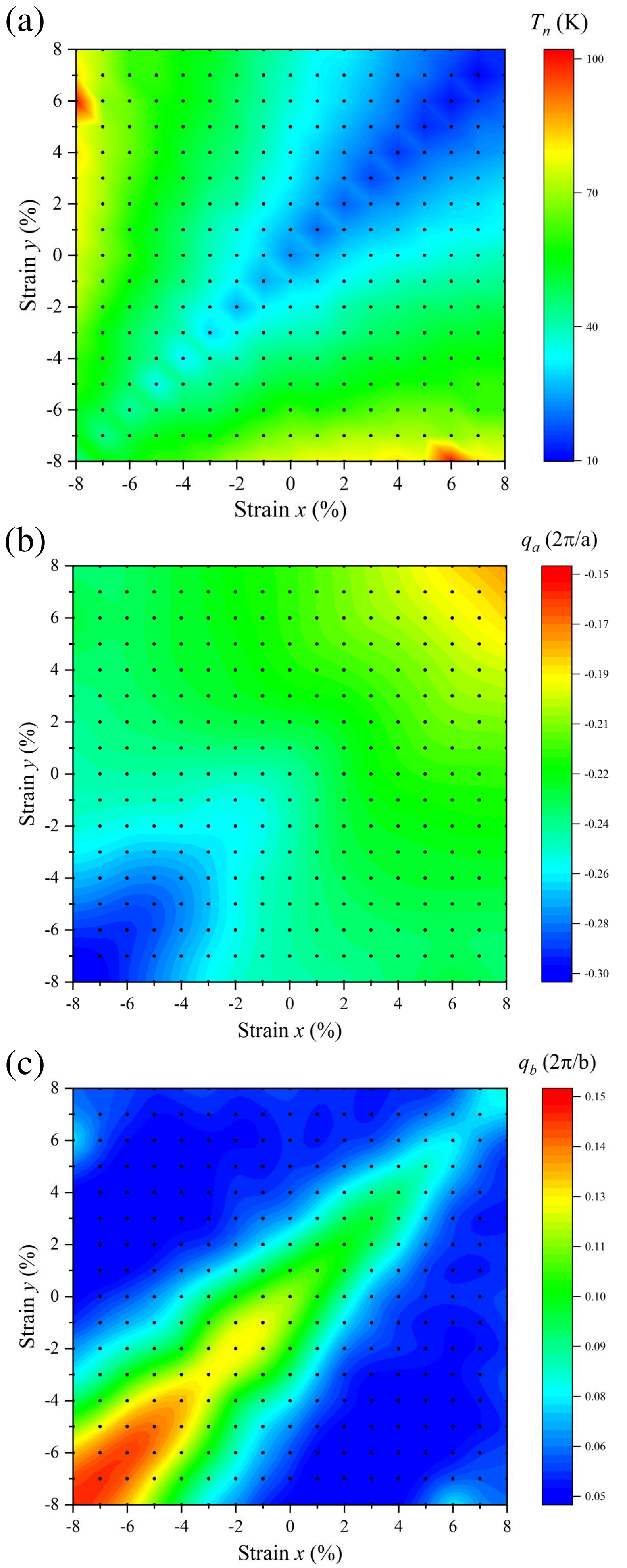}
	\caption{ (a) $T_{N}$ as a function of lattice constants $a$ and $b$, where the dark spots represent sampled strain points. (b)-(c) The $q_a$ and $q_b$ component of HM spin vector $\boldsymbol{Q}$ as a function of lattice constants $a$ and $b$.}
	\label{tu4}
\end{figure}

Our calculated $J$'s fulfill the requirement of Eq. (\ref{eq:6}), suggesting that the ground state of monolayer NiI$_{2}$ should be HM. Evidently, frustration from the strong FM $J_{1}$ and the AFM $J_{3}$ are the main driving force for the formation of the HM state, while $J_2$ is too weak to qualitatively contribute. MC simulations with the calculated parameters are then performed, confirming the presence of a single phase transtion from a paramagnetic phase to a HM phase, with a $T_{N}$ = 22.81 K, in good agreement with the experimental result of 21 K~\cite{9}. The HM state possesses a single $\boldsymbol{Q}$ , which has six symmetrically equivalent values(see Fig.~\ref{tu2}(a)), confirmed by Fourier analysis on the spin configurations output by the MC simulations. All six $\boldsymbol{Q}$ are along $C_{2}$ axis of monolayer NiI$_{2}$, i.e. [100], [010] and [110] directions, with $|\boldsymbol{Q}|$= 0.216 (2$\pi$)/$a$ (see Fig.~\ref{tu2}(a)). The two-component $\boldsymbol{Q}$'s are distinct from that used in literatures, where the in-plane component (0.138, 0, 0) of the experimental $\boldsymbol{Q}$ of the bulk NiI$_{2}$ is simply assumed~\cite{19}. To double check our model against DFT calculations, we further performed calculations on cycloid magnetic structures using the generalized Bloch theorem. The results show that the total energy of our HM state is lower than the most stable single-component $\boldsymbol{Q}$ state.

The electric polarization $\boldsymbol{P}$ of monolayer NiI$_{2}$ was reported to be parallel to the $C_2$ axis~\cite{9}. Given our calculated $\boldsymbol{Q}$ values, which are also parallel to the $C_2$ axis (see also illustration in Fig.~\ref{tu2}(a)), the measured $\boldsymbol{P}$ is expected to correspond to proper-screw HM states, according to symmetry arguements and calculations in isostructural MnI$_2$~\cite{18}. However, our model can not distinguish a cycloid from a proper-screw order (see Fig.~\ref{tu2}(b)). We thus again resort to direct DFT calculations. Without losing generality, we focus on the case with $\boldsymbol{Q}_{1}$ = (0.25, -0.125, 0), which can be simulated using a 4$\times$8$\times$1 supercell for both cycloid and proper-screw HM case. DFT calculations show that the proper-screw HM state is energetically more stable with 11.67 meV lower than the cycloidal one. We therefore propose that the magnetic ground state of monolayer NiI$_{2}$ is a proper-screw HM state. It is worth noting that the I atoms are also calculated to develop sizable magnetizations about 0.20 $\mu_B$, forming a proper-screw HM state in coordination with that formed by the Ni spins(see Fig.~\ref{tu2}(c)). Furthermore for ferroelectricity, we obtain an estimation of the polarization with an amplidute $|\boldsymbol{P}|$ $\sim$ 2.62 $\times$ 10$^{-13}$ Cm$^{-1}$, with directions parallel to each $\boldsymbol{Q}$ (see Fig.~\ref{tu2}(a)). 


We further systematically investigate the electronic structure of monolayer NiI$_{2}$. For comparison, the band structure of the FM state without SOC is first calculated. As shown in Fig.~\ref{tu2}(d), the band gap is 3.13 eV for the spin-up channel and 1.27 eV for the spin-down channel, respectively. This significant difference between the two band gaps indicates that the FM monolayer NiI$_{2}$ is a half-semiconductor~\cite{AN2022115262,dong2022quantum}, which has great potential for spintronic applications~\cite{35}. The conduction band minimum (CBM) is composed of spin-down states only, while the valence band maximum (VBM) is composed of spin-up states, leading to an overall band gap of 1.05 eV in the FM monolayer NiI$_{2}$. However, the band gap in the FM state with SOC decreases to 0.73 eV. We further use a 4$\times$8$\times$1 supercell to calculate the HM state with $\boldsymbol{Q}_{1}$ = (0.25, -0.125, 0), with the bands unfolded to the Brillouin zone of a primitive cell~\cite{PhysRevLett.104.216401}. As shown in  Fig.~\ref{tu2}(e), without SOC, the monolayer NiI$_{2}$ with the HM state is calculated to have a 1.23 eV indirect band gap near the $\Gamma$ point. When SOC is considered, the  CBM moves downward, leading to a slightly smaller gap (Fig.~\ref{tu2}(f)). This overall semiconducting nature of the HM state provides a prerequisite for the appearance of spontaneous electric polarization in the monolayer NiI$_{2}$.

\subsection{Strain engineered monolayer NiI$_{2}$}

We firstly investigate the case of applying uniform biaxial strain, i.e., the same strain on both the $a$-axis and $b$-aixs, to gain an insight about how strain generally affect the atomic structure and exchange interactions, hence the magnetic ordering. In this case, the original lattice symmetry is preserved. With the strain changed from -8\% to 8 \%, the bond lengths of $J_{1}$, $J_{2}$ and $J_{3}$ extend from 3.66 \AA\ to 4.30 \AA,  6.34 \AA\ to 7.44 \AA, 7.32 \AA\ to 8.59 \AA\ , respectively. Fig.~\ref{tu3}(a) shows the evolvement of the magnetic exchange interactions with the strain level $x$. It can be seen that $J_{1}$ keeps ferromagnetic, showing only slight variations. In contrast, although the $J_{3}$ remains antiferromagnetic, it decreases dramatically and monotonously with the increase of $x$, exhibiting much more tunability. Meanwhile, $J_{2}$ remains negligiblly close to 0 throughout the whole range of $x$, therefore Eq. (\ref{eq:6}) can be simplified as
 \begin{equation}
\begin{aligned}
\frac{J_1}{J_3} > -4
\end{aligned}\label{eq:7}
\end{equation}
Our calculated ratio of $J_{1}$ and $J_{3}$ ranges from -2.25 to -0.24, satisfying Eq. (\ref{eq:7}), therefore suggests that the ground states remain HM under these strains.

To further explore the impact of different kind of strains on magnetic interactions, we now consider the case of uniaxial strains, i.e., zero strain on the $a$-axis combined with varying strain levels $y$ on the $b$-axis. The $J_{1}$, $J_{2}$ and $J_{3}$ now split into  $J_{1a-1c}$ , $J_{2a-2c}$ and $J_{3a-3c}$ respectively, due to a reduced lattice symmetry. In Fig.~\ref{tu3}(b), it can be seen that $J_{1a-1c}$ and $J_{2a-2c}$ are overall not senstitive to the strains, similar to the behavior of their counterparts in the uniform strain case.  Exceptionally, $J_{3a-3c}$ behave differently among each other, with the $J_{3a}$ showing signicantly higher tunability, inheriting the characteristic of $J_3$. Particularly, under maximum tensile strain with $y = -8 \%$, although only small variation in the $J_{3a-3c}$ bond length is induced, $J_{3a}$ becomes over 4 times larger than $J_{3b}$ (see Fig.~\ref{tu3}(b)).

\begin{figure}[t]
	\includegraphics[scale=0.5]{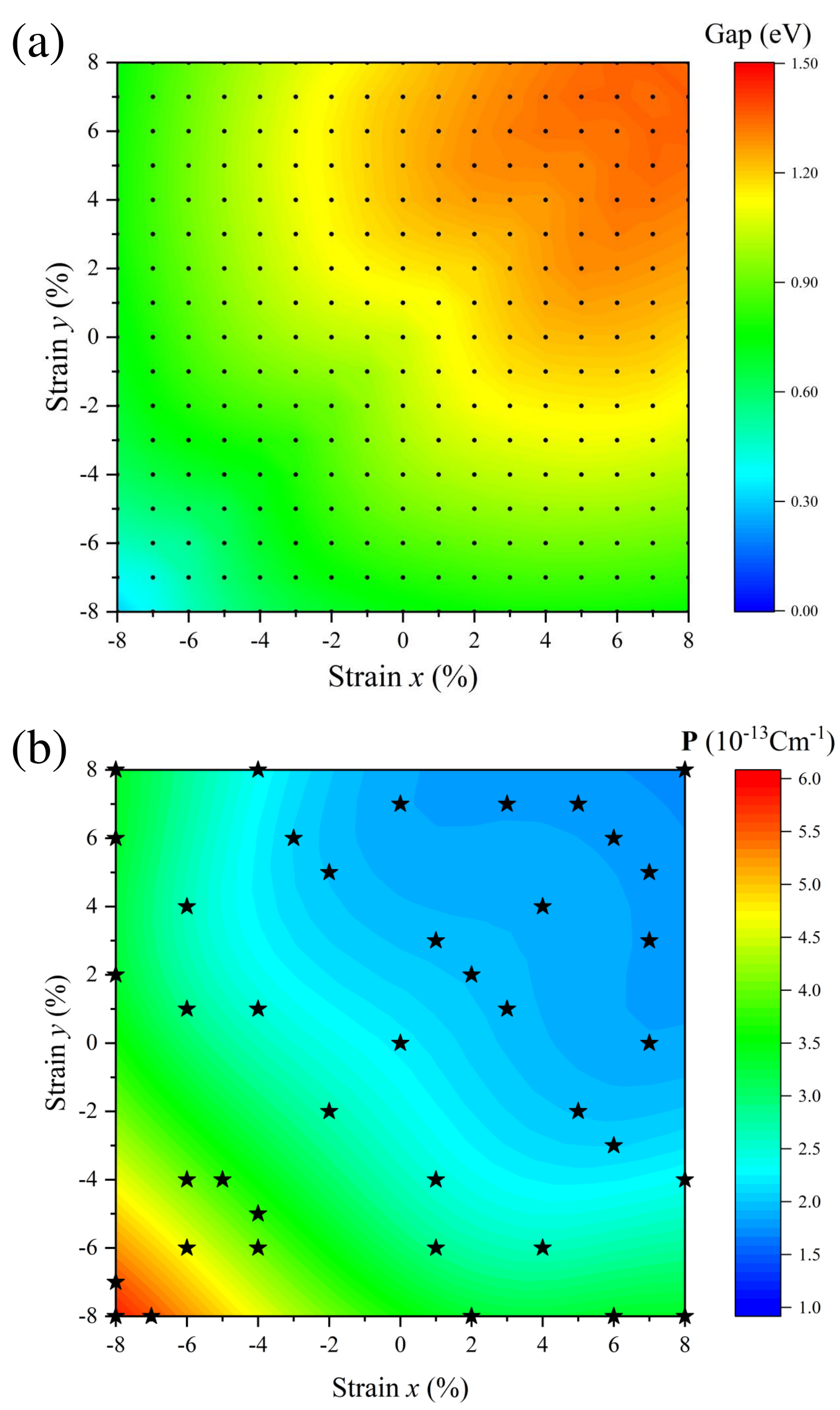}
	\caption{ (a) Band gap of the FM NiI$_{2}$ monolayer as a function of lattice constants $a$ and $b$, where the dark spots represent sampled strain points. (b) The polarization of monolayer NiI$_{2}$ in proper-screw HM states as a function of lattice constants $a$ and $b$, where the the dark stars are data points form the supercell calculation.}
	\label{tu5}
\end{figure}

We now turn to arbitrary strains with unequal $x$ and $y$ ( -8\% $\leq$ $x$, $y$ $\leq$ 8\% ). We find that in all scenarios, the calculated $J's$ behave similarly as been shown in Fig.~\ref{tu3}(a)-(b), with only $J_3$ varies significantly with $x$ and $y$. Importantly, the magnetic ground states are all predicted to be HM according to the Eq. (\ref{eq:7}). With the calculated $J'$s, we further perform MC simulations for all sampled strains, with calculated $T_{N}$ shown in Fig.~\ref{tu4}(a). The corresponding color map is plotted to be symmetrical with respect to the diagonal line since the $a$-axis and $b$-axis are equivalent. Notably, the $T_{N}$ exhibits dramatic variations, dropping with the increasing of tensile strains, down to 10.45 K at $x=8\%$ and $y=8\%$. Surprisingly, the $T_{N}$ rises to 101.17 K with the $x=-8 \%$\ and $y = 6\%$, which is the highest among all our sampled cases, almost 5 times higher than that of the strain-free one. By examining the corresponding $J'$s, we find that the $T_{N}$ largely evolves following the trend of the $J_{3}$, and in the case of lowered symmetry, $J_{3a}$, which is the exchange interaction with the highest tunability (see Fig.~\ref{tu3}(a)-(b) and Table.~\ref{tab1-1} for typical strain cases). We then further identify the HM propogation vectors $\boldsymbol{Q}$ from the outputed spin configurations of MC simulations, with $q_a$ and $q_b$ shown in Fig.~\ref{tu4}(b)-\ref{tu4}(c) respectively. It can be seen that the amplitude of $q_a$ are enlarged with the increasing tensile strains, while the $q_b$ increases with the compressive strains (the diagonal lines of the color maps). 

\begin{figure}[t]
	\includegraphics[scale=0.45]{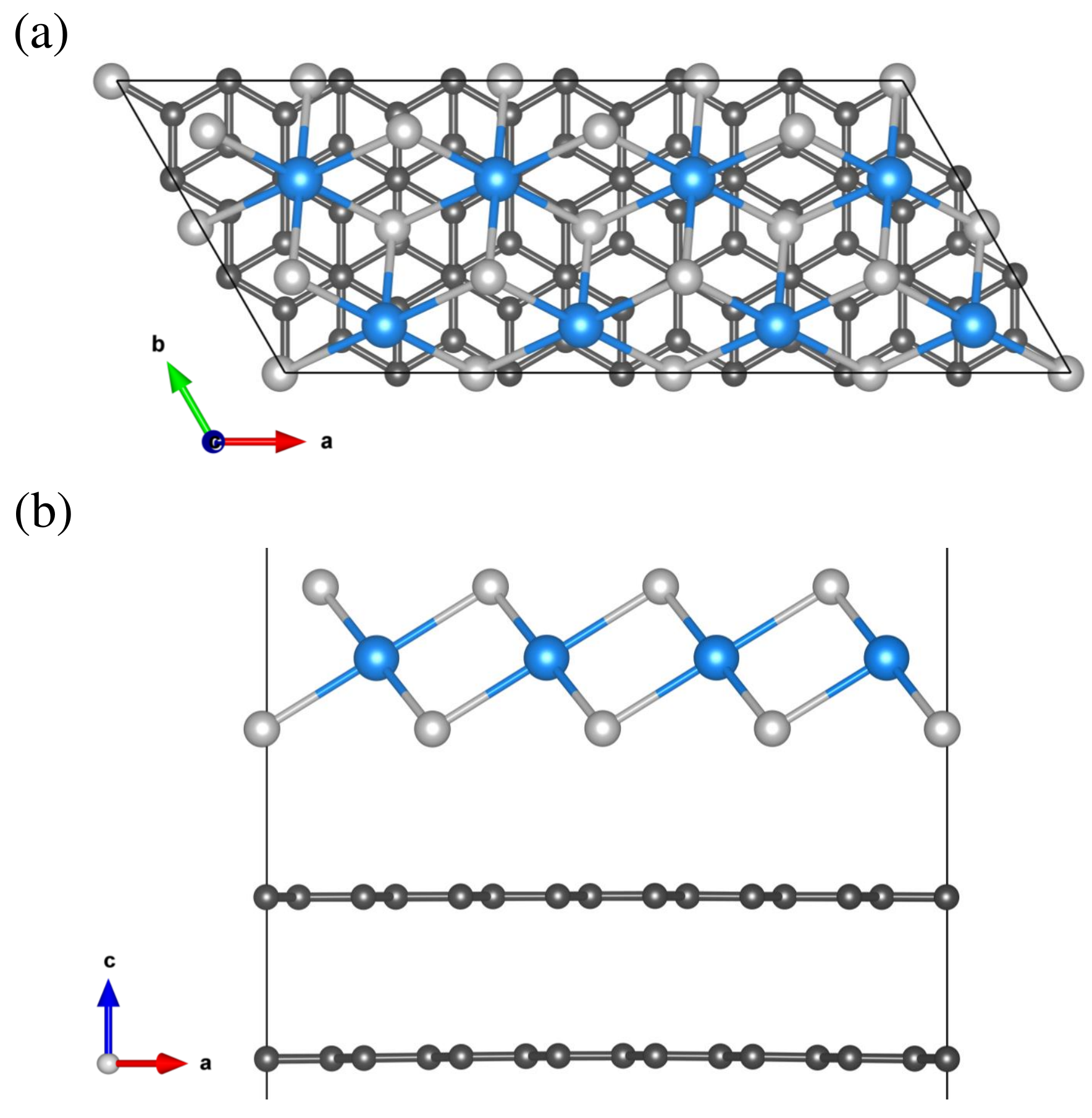}
	\caption{ (a)-(b) The top and side view of  4 $\times$ 2 $\times$ 1 NiI$_{2}$ (monolayer)/ 7 $\times$ 3 $\times$ 1 graphene (bilayer) heterostructures.}
	\label{tu6}
\end{figure}

\begin{table*}[t!]
	\caption{ \label{tab2} Values (in meV) of magnetic exchange couplings and $T_{N}$ by DFT calculation and MC simulations.The freestanding monolayer NiI$_{2}$ is under $x$ = -7\% and $y$ = 8\% strain, where the NiI$_{2}$ layer in NiI$_{2}$/BLG heterostructure is taking similar compressive and tensile strains. }
	\begin{ruledtabular}
		\begin{tabular}{cccccccccccc}
			&  & $ J_{1a}$ & $ J_{1b}$ & $ J_{1c}$ & $ J_{2a}$  & $ J_{2b}$ & $ J_{2c}$ & $ J_{3a}$ & $ J_{3b}$ & $ J_3c$ & $T_{N}$(K) \\
			\hline  \\[-1.0ex]
			& NiI$_{2}$ monolayer & -2.98 & -4.18  & -4.58 & 0.08 & -0.15 & -0.20 & 12.38 & 2.49 & 1.02 & 71.07  \\
			& NiI$_{2}$/BLG heterostructure & -3.50 & -4.27  & -4.51 & 0.22 & -0.09 & -0.37 & 12.06 & 2.81 & 1.22 & 70.73  \\
		\end{tabular}
	\end{ruledtabular}
\end{table*}

Once the $\boldsymbol{Q}$ values are confirmed, we construct supercells to accommodate corresponding proper-screw spiral states to study their ferroelectricity. Since the supercell calculations are expensive, to reduce computational effort, we only perform supercell calculations on selected representative strain values, shown as stars in Fig.~\ref{tu5}(b). We find that all the selected HM states have larger band gaps than that of the corresponding FM states. Therefore, to have a full sampling, band gaps of FM states with all strains are calculated and shown as a color map in Fig.~\ref{tu5}(a). It can be seen that in all cases, the band gaps are finite, ranging from 0.30 eV to 1.36 eV, which increase with tensile strains and reduce with compressive strains, consistent with the general text-book physical picture. These band gaps guarantee that the HM ground states under the whole strain range are possible to sustain spontaneous electric polarizations. The electric polarziations $\boldsymbol{P}$ are then calculated, with their amplitude $|\boldsymbol{P}|$ interpolated to a diagram in the strain map. As can be seen in Fig.~\ref{tu5}(b), $|\boldsymbol{P}|$ keeps the same order in the whole strain range, with their behavior closely resembling that of the band gaps, but in an opposite way, which is quite evident from the comparison of Fig.~\ref{tu5}(a) and Fig.~\ref{tu5}(b). Therefore, the main driving force for the enhancement of  $|\boldsymbol{P}|$ is the corresponding shrinking of the bang gap, although with compressive strains, the surface of the monolayer NiI$_{2}$ also contracts. For instance, with the maximum compressive strain $x = -8 \%$ and $y = -8 \%$, the surface area of the monolayer NiI$_{2}$ is reduced by 15 \%, but $|\boldsymbol{P}|$ becomes more than two times larger than that of the strain-free case.

To check the reliability of applying strain by solely changing the lattice constants, we construct a NiI$_{2}$/graphene heterostructure to simulate practical exitaxial strain engineering through substate.  Bilayer graphene (BLG) has been successfully used as substrates to grow 2D magnetic materials, such as monolayer CrTe$_{2}$~\cite{41}. As shown in Fig.~\ref{tu6}(a), the constructed NiI$_{2}$/BLG heterostructure is composed of a 4 $\times$ 2 $\times$ 1 supercell of NiI$_{2}$ and a 7 $\times$ 3 $\times$ 1 supercell of BLG. By fixing the lattice constants of the BLG, 7.24\% compressive strain ($x=-7.24\%$) and 8.22\% tensile strain ($y=8.22\%$) are applied on the $a$-axis and $b$-axis of monolayer NiI$_{2}$ , respectively. The calculated magnetic exchange interactions $J$$_{1a-1c}$ , $J$$_{2a-2c}$ and $J$$_{3a-3c}$ in this heterostructure are shown in Table.~\ref{tab2}, resulting in a HM ground state with a $T$$_{N}$ = 71.07 K. In comparison, as shown in Fig.~\ref{tu4}(a), our calculations of freestanding monolayer NiI$_{2}$ with strain values of $x = -7\%$ and $y = 8\%$ produces a $T$$_{N}$ = 70.73 K (Table.~\ref{tab2}), with the corresponding $J$ values very close to that from the heterostructures. This excellent agreement validates the high reliablity of our predictions based on freestanding calculations.

\section{Summary}
In summary, we study the multiferroic order of freestanding monolayer NiI$_{2}$ with engineered levels of strain using DFT calculations and MC simulations. Through the investigation of strain-free monolayer NiI$_{2}$, we find that the first NN and third NN exchange interactions play an essential role in the formation of its magnetic phase diagrams. The competition of these interactions induces magnetic frustration, leading to the formation of proper-screw HM ground state. We further show that these conclusions drawing from the strain-free monolayer can be well generalized to the cases within our engineered range of strains. Notably, our calculations show that with 6\% tensile strain on the $a$-axis and 8\% compressive strain on the $b$-axis, the N\'{e}el temperature $T_N$ can be significantly enhanced to 101 K, about 5 times larger than that of the strain-free one. The strength of spontaneous electric polarizations can also be more than doubled under 8\% uniform compressive strain on both axises. Moreover, we construct a NiI$_{2}$ (monolayer)/Graphene (bilayer) heterostructure to simulate a more realistic strain engineering and obtain results in good agreement with that based on calculations on corresponding freestanding one. Our work suggests that strain is a promising way to tune multiferroic orders in the monolayer NiI$_{2}$, with the potential to signicanlty promote its transition temperatures and electric polarizations, therefore broaden the prospect of its applications in spintronics devices.

\section{acknowledgments}

Work at Sun Yat-Sen University was supported by the National Key Research and Development Program of China (Grants No. 2018YFA0306001, 2017YFA0206203), and the Guangdong Basic and Applied Basic Research Foundation (Grants No. 2022A1515011618, No. 2019A1515011337), and the National Natural Science Foundation of China (Grants No. 92165204, No. 11974432), and the Shenzhen International Quantum Academy (Grant No. SIQA202102), Leading Talent Program of Guangdong Special Projects (201626003).


\begin{thebibliography}{51}%
\makeatletter
\providecommand \@ifxundefined [1]{%
 \@ifx{#1\undefined}
}%
\providecommand \@ifnum [1]{%
 \ifnum #1\expandafter \@firstoftwo
 \else \expandafter \@secondoftwo
 \fi
}%
\providecommand \@ifx [1]{%
 \ifx #1\expandafter \@firstoftwo
 \else \expandafter \@secondoftwo
 \fi
}%
\providecommand \natexlab [1]{#1}%
\providecommand \enquote  [1]{``#1''}%
\providecommand \bibnamefont  [1]{#1}%
\providecommand \bibfnamefont [1]{#1}%
\providecommand \citenamefont [1]{#1}%
\providecommand \href@noop [0]{\@secondoftwo}%
\providecommand \href [0]{\begingroup \@sanitize@url \@href}%
\providecommand \@href[1]{\@@startlink{#1}\@@href}%
\providecommand \@@href[1]{\endgroup#1\@@endlink}%
\providecommand \@sanitize@url [0]{\catcode `\\12\catcode `\$12\catcode
  `\&12\catcode `\#12\catcode `\^12\catcode `\_12\catcode `\%12\relax}%
\providecommand \@@startlink[1]{}%
\providecommand \@@endlink[0]{}%
\providecommand \url  [0]{\begingroup\@sanitize@url \@url }%
\providecommand \@url [1]{\endgroup\@href {#1}{\urlprefix }}%
\providecommand \urlprefix  [0]{URL }%
\providecommand \Eprint [0]{\href }%
\providecommand \doibase [0]{https://doi.org/}%
\providecommand \selectlanguage [0]{\@gobble}%
\providecommand \bibinfo  [0]{\@secondoftwo}%
\providecommand \bibfield  [0]{\@secondoftwo}%
\providecommand \translation [1]{[#1]}%
\providecommand \BibitemOpen [0]{}%
\providecommand \bibitemStop [0]{}%
\providecommand \bibitemNoStop [0]{.\EOS\space}%
\providecommand \EOS [0]{\spacefactor3000\relax}%
\providecommand \BibitemShut  [1]{\csname bibitem#1\endcsname}%
\let\auto@bib@innerbib\@empty
\bibitem [{\citenamefont {Jariwala}\ \emph {et~al.}(2017)\citenamefont
  {Jariwala}, \citenamefont {Davoyan}, \citenamefont {Wong},\ and\
  \citenamefont {Atwater}}]{1}%
  \BibitemOpen
  \bibfield  {author} {\bibinfo {author} {\bibfnamefont {D.}~\bibnamefont
  {Jariwala}}, \bibinfo {author} {\bibfnamefont {A.~R.}\ \bibnamefont
  {Davoyan}}, \bibinfo {author} {\bibfnamefont {J.}~\bibnamefont {Wong}},\ and\
  \bibinfo {author} {\bibfnamefont {H.~A.}\ \bibnamefont {Atwater}},\
  }\bibfield  {title} {\bibinfo {title} {{Van der Waals materials for
  atomically-thin photovoltaics: promise and outlook}},\ }\href
  {https://doi.org/10.1021/acsphotonics.7b01103} {\bibfield  {journal}
  {\bibinfo  {journal} {Acs Photonics}\ }\textbf {\bibinfo {volume} {4}},\
  \bibinfo {pages} {2962} (\bibinfo {year} {2017})}\BibitemShut {NoStop}%
\bibitem [{\citenamefont {Huang}\ \emph {et~al.}(2017)\citenamefont {Huang},
  \citenamefont {Clark}, \citenamefont {Navarro-Moratalla}, \citenamefont
  {Klein}, \citenamefont {Cheng}, \citenamefont {Seyler}, \citenamefont
  {Zhong}, \citenamefont {Schmidgall}, \citenamefont {McGuire}, \citenamefont
  {Cobden} \emph {et~al.}}]{2}%
  \BibitemOpen
  \bibfield  {author} {\bibinfo {author} {\bibfnamefont {B.}~\bibnamefont
  {Huang}}, \bibinfo {author} {\bibfnamefont {G.}~\bibnamefont {Clark}},
  \bibinfo {author} {\bibfnamefont {E.}~\bibnamefont {Navarro-Moratalla}},
  \bibinfo {author} {\bibfnamefont {D.~R.}\ \bibnamefont {Klein}}, \bibinfo
  {author} {\bibfnamefont {R.}~\bibnamefont {Cheng}}, \bibinfo {author}
  {\bibfnamefont {K.~L.}\ \bibnamefont {Seyler}}, \bibinfo {author}
  {\bibfnamefont {D.}~\bibnamefont {Zhong}}, \bibinfo {author} {\bibfnamefont
  {E.}~\bibnamefont {Schmidgall}}, \bibinfo {author} {\bibfnamefont {M.~A.}\
  \bibnamefont {McGuire}}, \bibinfo {author} {\bibfnamefont {D.~H.}\
  \bibnamefont {Cobden}}, \emph {et~al.},\ }\bibfield  {title} {\bibinfo
  {title} {{Layer-dependent ferromagnetism in a van der Waals crystal down to
  the monolayer limit}},\ }\href {https://doi.org/10.1038/nature22391}
  {\bibfield  {journal} {\bibinfo  {journal} {Nature}\ }\textbf {\bibinfo
  {volume} {546}},\ \bibinfo {pages} {270} (\bibinfo {year}
  {2017})}\BibitemShut {NoStop}%
\bibitem [{\citenamefont {Bian}\ \emph {et~al.}(2022)\citenamefont {Bian},
  \citenamefont {Li}, \citenamefont {Liu}, \citenamefont {Cao}, \citenamefont
  {Fu}, \citenamefont {Meng}, \citenamefont {Zhou}, \citenamefont {Liu},\ and\
  \citenamefont {Liu}}]{2-1}%
  \BibitemOpen
  \bibfield  {author} {\bibinfo {author} {\bibfnamefont {R.}~\bibnamefont
  {Bian}}, \bibinfo {author} {\bibfnamefont {C.}~\bibnamefont {Li}}, \bibinfo
  {author} {\bibfnamefont {Q.}~\bibnamefont {Liu}}, \bibinfo {author}
  {\bibfnamefont {G.}~\bibnamefont {Cao}}, \bibinfo {author} {\bibfnamefont
  {Q.}~\bibnamefont {Fu}}, \bibinfo {author} {\bibfnamefont {P.}~\bibnamefont
  {Meng}}, \bibinfo {author} {\bibfnamefont {J.}~\bibnamefont {Zhou}}, \bibinfo
  {author} {\bibfnamefont {F.}~\bibnamefont {Liu}},\ and\ \bibinfo {author}
  {\bibfnamefont {Z.}~\bibnamefont {Liu}},\ }\bibfield  {title} {\bibinfo
  {title} {{Recent progress in the synthesis of novel two-dimensional van der
  Waals materials}},\ }\href {https://doi.org/10.1093/nsr/nwab164} {\bibfield
  {journal} {\bibinfo  {journal} {National Science Review}\ }\textbf {\bibinfo
  {volume} {9}},\ \bibinfo {pages} {nwab164} (\bibinfo {year}
  {2022})}\BibitemShut {NoStop}%
\bibitem [{\citenamefont {Deng}\ \emph {et~al.}(2020)\citenamefont {Deng},
  \citenamefont {Yu}, \citenamefont {Shi}, \citenamefont {Guo}, \citenamefont
  {Xu}, \citenamefont {Wang}, \citenamefont {Chen},\ and\ \citenamefont
  {Zhang}}]{2-2}%
  \BibitemOpen
  \bibfield  {author} {\bibinfo {author} {\bibfnamefont {Y.}~\bibnamefont
  {Deng}}, \bibinfo {author} {\bibfnamefont {Y.}~\bibnamefont {Yu}}, \bibinfo
  {author} {\bibfnamefont {M.~Z.}\ \bibnamefont {Shi}}, \bibinfo {author}
  {\bibfnamefont {Z.}~\bibnamefont {Guo}}, \bibinfo {author} {\bibfnamefont
  {Z.}~\bibnamefont {Xu}}, \bibinfo {author} {\bibfnamefont {J.}~\bibnamefont
  {Wang}}, \bibinfo {author} {\bibfnamefont {X.~H.}\ \bibnamefont {Chen}},\
  and\ \bibinfo {author} {\bibfnamefont {Y.}~\bibnamefont {Zhang}},\ }\bibfield
   {title} {\bibinfo {title} {{Quantum anomalous Hall effect in intrinsic
  magnetic topological insulator MnBi$_2$Te$_4$}},\ }\href
  {https://doi.org/10.1126/science.aax8156} {\bibfield  {journal} {\bibinfo
  {journal} {Science}\ }\textbf {\bibinfo {volume} {367}},\ \bibinfo {pages}
  {895} (\bibinfo {year} {2020})}\BibitemShut {NoStop}%
\bibitem [{\citenamefont {Wu}\ \emph {et~al.}(2019)\citenamefont {Wu},
  \citenamefont {Liu}, \citenamefont {Sasase}, \citenamefont {Ienaga},
  \citenamefont {Obata}, \citenamefont {Yukawa}, \citenamefont {Horiba},
  \citenamefont {Kumigashira}, \citenamefont {Okuma}, \citenamefont {Inoshita}
  \emph {et~al.}}]{2-3}%
  \BibitemOpen
  \bibfield  {author} {\bibinfo {author} {\bibfnamefont {J.}~\bibnamefont
  {Wu}}, \bibinfo {author} {\bibfnamefont {F.}~\bibnamefont {Liu}}, \bibinfo
  {author} {\bibfnamefont {M.}~\bibnamefont {Sasase}}, \bibinfo {author}
  {\bibfnamefont {K.}~\bibnamefont {Ienaga}}, \bibinfo {author} {\bibfnamefont
  {Y.}~\bibnamefont {Obata}}, \bibinfo {author} {\bibfnamefont
  {R.}~\bibnamefont {Yukawa}}, \bibinfo {author} {\bibfnamefont
  {K.}~\bibnamefont {Horiba}}, \bibinfo {author} {\bibfnamefont
  {H.}~\bibnamefont {Kumigashira}}, \bibinfo {author} {\bibfnamefont
  {S.}~\bibnamefont {Okuma}}, \bibinfo {author} {\bibfnamefont
  {T.}~\bibnamefont {Inoshita}}, \emph {et~al.},\ }\bibfield  {title} {\bibinfo
  {title} {{Natural van der Waals heterostructural single crystals with both
  magnetic and topological properties}},\ }\href
  {https://doi.org/10.1126/sciadv.aax9989} {\bibfield  {journal} {\bibinfo
  {journal} {Science advances}\ }\textbf {\bibinfo {volume} {5}},\ \bibinfo
  {pages} {eaax9989} (\bibinfo {year} {2019})}\BibitemShut {NoStop}%
\bibitem [{\citenamefont {Li}\ \emph {et~al.}(2021)\citenamefont {Li},
  \citenamefont {Song}, \citenamefont {Zhao}, \citenamefont {Vaklinova},
  \citenamefont {Zhao}, \citenamefont {Li}, \citenamefont {Qiu}, \citenamefont
  {Wang}, \citenamefont {Lin}, \citenamefont {Zhao} \emph {et~al.}}]{2-4}%
  \BibitemOpen
  \bibfield  {author} {\bibinfo {author} {\bibfnamefont {J.}~\bibnamefont
  {Li}}, \bibinfo {author} {\bibfnamefont {P.}~\bibnamefont {Song}}, \bibinfo
  {author} {\bibfnamefont {J.}~\bibnamefont {Zhao}}, \bibinfo {author}
  {\bibfnamefont {K.}~\bibnamefont {Vaklinova}}, \bibinfo {author}
  {\bibfnamefont {X.}~\bibnamefont {Zhao}}, \bibinfo {author} {\bibfnamefont
  {Z.}~\bibnamefont {Li}}, \bibinfo {author} {\bibfnamefont {Z.}~\bibnamefont
  {Qiu}}, \bibinfo {author} {\bibfnamefont {Z.}~\bibnamefont {Wang}}, \bibinfo
  {author} {\bibfnamefont {L.}~\bibnamefont {Lin}}, \bibinfo {author}
  {\bibfnamefont {M.}~\bibnamefont {Zhao}}, \emph {et~al.},\ }\bibfield
  {title} {\bibinfo {title} {Printable two-dimensional superconducting
  monolayers},\ }\href {https://doi.org/10.1038/s41563-020-00831-1} {\bibfield
  {journal} {\bibinfo  {journal} {Nature Materials}\ }\textbf {\bibinfo
  {volume} {20}},\ \bibinfo {pages} {181} (\bibinfo {year} {2021})}\BibitemShut
  {NoStop}%
\bibitem [{\citenamefont {Burch}\ \emph {et~al.}(2018)\citenamefont {Burch},
  \citenamefont {Mandrus},\ and\ \citenamefont {Park}}]{5}%
  \BibitemOpen
  \bibfield  {author} {\bibinfo {author} {\bibfnamefont {K.~S.}\ \bibnamefont
  {Burch}}, \bibinfo {author} {\bibfnamefont {D.}~\bibnamefont {Mandrus}},\
  and\ \bibinfo {author} {\bibfnamefont {J.-G.}\ \bibnamefont {Park}},\
  }\bibfield  {title} {\bibinfo {title} {{Magnetism in two-dimensional van der
  Waals materials}},\ }\href {https://doi.org/10.1038/s41586-018-0631-z}
  {\bibfield  {journal} {\bibinfo  {journal} {Nature}\ }\textbf {\bibinfo
  {volume} {563}},\ \bibinfo {pages} {47} (\bibinfo {year} {2018})}\BibitemShut
  {NoStop}%
\bibitem [{\citenamefont {Park}(2016)}]{6}%
  \BibitemOpen
  \bibfield  {author} {\bibinfo {author} {\bibfnamefont {J.-G.}\ \bibnamefont
  {Park}},\ }\bibfield  {title} {\bibinfo {title} {{Opportunities and
  challenges of 2D magnetic van der Waals materials: magnetic graphene?}},\
  }\href {https://doi.org/10.1088/0953-8984/28/30/301001} {\bibfield  {journal}
  {\bibinfo  {journal} {Journal of Physics: Condensed Matter}\ }\textbf
  {\bibinfo {volume} {28}},\ \bibinfo {pages} {301001} (\bibinfo {year}
  {2016})}\BibitemShut {NoStop}%
\bibitem [{\citenamefont {Bonilla}\ \emph {et~al.}(2018)\citenamefont
  {Bonilla}, \citenamefont {Kolekar}, \citenamefont {Ma}, \citenamefont {Diaz},
  \citenamefont {Kalappattil}, \citenamefont {Das}, \citenamefont {Eggers},
  \citenamefont {Gutierrez}, \citenamefont {Phan},\ and\ \citenamefont
  {Batzill}}]{6-1}%
  \BibitemOpen
  \bibfield  {author} {\bibinfo {author} {\bibfnamefont {M.}~\bibnamefont
  {Bonilla}}, \bibinfo {author} {\bibfnamefont {S.}~\bibnamefont {Kolekar}},
  \bibinfo {author} {\bibfnamefont {Y.}~\bibnamefont {Ma}}, \bibinfo {author}
  {\bibfnamefont {H.~C.}\ \bibnamefont {Diaz}}, \bibinfo {author}
  {\bibfnamefont {V.}~\bibnamefont {Kalappattil}}, \bibinfo {author}
  {\bibfnamefont {R.}~\bibnamefont {Das}}, \bibinfo {author} {\bibfnamefont
  {T.}~\bibnamefont {Eggers}}, \bibinfo {author} {\bibfnamefont {H.~R.}\
  \bibnamefont {Gutierrez}}, \bibinfo {author} {\bibfnamefont {M.-H.}\
  \bibnamefont {Phan}},\ and\ \bibinfo {author} {\bibfnamefont
  {M.}~\bibnamefont {Batzill}},\ }\bibfield  {title} {\bibinfo {title} {{Strong
  room-temperature ferromagnetism in VSe$_{2}$ monolayers on van der Waals
  substrates}},\ }\href {https://doi.org/10.1038/s41565-018-0063-9} {\bibfield
  {journal} {\bibinfo  {journal} {Nature nanotechnology}\ }\textbf {\bibinfo
  {volume} {13}},\ \bibinfo {pages} {289} (\bibinfo {year} {2018})}\BibitemShut
  {NoStop}%
\bibitem [{\citenamefont {Lin}\ and\ \citenamefont {Chen}(2020)}]{6-2}%
  \BibitemOpen
  \bibfield  {author} {\bibinfo {author} {\bibfnamefont {Z.-Z.}\ \bibnamefont
  {Lin}}\ and\ \bibinfo {author} {\bibfnamefont {X.}~\bibnamefont {Chen}},\
  }\bibfield  {title} {\bibinfo {title} {{Ultrathin scattering spin filter and
  magnetic tunnel junction implemented by ferromagnetic 2D van der Waals
  material}},\ }\href {https://doi.org/10.1002/aelm.201900968} {\bibfield
  {journal} {\bibinfo  {journal} {Advanced Electronic Materials}\ }\textbf
  {\bibinfo {volume} {6}},\ \bibinfo {pages} {1900968} (\bibinfo {year}
  {2020})}\BibitemShut {NoStop}%
\bibitem [{\citenamefont {Park}\ \emph {et~al.}(1999)\citenamefont {Park},
  \citenamefont {Kang}, \citenamefont {Bu}, \citenamefont {Noh}, \citenamefont
  {Lee},\ and\ \citenamefont {Jo}}]{7-1}%
  \BibitemOpen
  \bibfield  {author} {\bibinfo {author} {\bibfnamefont {B.}~\bibnamefont
  {Park}}, \bibinfo {author} {\bibfnamefont {B.}~\bibnamefont {Kang}}, \bibinfo
  {author} {\bibfnamefont {S.}~\bibnamefont {Bu}}, \bibinfo {author}
  {\bibfnamefont {T.}~\bibnamefont {Noh}}, \bibinfo {author} {\bibfnamefont
  {J.}~\bibnamefont {Lee}},\ and\ \bibinfo {author} {\bibfnamefont
  {W.}~\bibnamefont {Jo}},\ }\bibfield  {title} {\bibinfo {title}
  {Lanthanum-substituted bismuth titanate for use in non-volatile memories},\
  }\href {https://doi.org/10.1038/44352} {\bibfield  {journal} {\bibinfo
  {journal} {Nature}\ }\textbf {\bibinfo {volume} {401}},\ \bibinfo {pages}
  {682} (\bibinfo {year} {1999})}\BibitemShut {NoStop}%
\bibitem [{\citenamefont {Scott}(2007)}]{7-2}%
  \BibitemOpen
  \bibfield  {author} {\bibinfo {author} {\bibfnamefont {J.}~\bibnamefont
  {Scott}},\ }\bibfield  {title} {\bibinfo {title} {Applications of modern
  ferroelectrics},\ }\href {https://doi.org/10.1126/science.112956} {\bibfield
  {journal} {\bibinfo  {journal} {science}\ }\textbf {\bibinfo {volume}
  {315}},\ \bibinfo {pages} {954} (\bibinfo {year} {2007})}\BibitemShut
  {NoStop}%
\bibitem [{\citenamefont {Han}\ \emph {et~al.}(2013)\citenamefont {Han},
  \citenamefont {Zhou},\ and\ \citenamefont {Roy}}]{7-3}%
  \BibitemOpen
  \bibfield  {author} {\bibinfo {author} {\bibfnamefont {S.-T.}\ \bibnamefont
  {Han}}, \bibinfo {author} {\bibfnamefont {Y.}~\bibnamefont {Zhou}},\ and\
  \bibinfo {author} {\bibfnamefont {V.}~\bibnamefont {Roy}},\ }\bibfield
  {title} {\bibinfo {title} {Towards the development of flexible non-volatile
  memories},\ }\href {https://doi.org/10.1002/adma.201301361} {\bibfield
  {journal} {\bibinfo  {journal} {Advanced Materials}\ }\textbf {\bibinfo
  {volume} {25}},\ \bibinfo {pages} {5425} (\bibinfo {year}
  {2013})}\BibitemShut {NoStop}%
\bibitem [{\citenamefont {Cui}\ \emph {et~al.}(2018)\citenamefont {Cui},
  \citenamefont {Hu}, \citenamefont {Yan}, \citenamefont {Addiego},
  \citenamefont {Gao}, \citenamefont {Wang}, \citenamefont {Wang},
  \citenamefont {Li}, \citenamefont {Cheng}, \citenamefont {Li} \emph
  {et~al.}}]{7}%
  \BibitemOpen
  \bibfield  {author} {\bibinfo {author} {\bibfnamefont {C.}~\bibnamefont
  {Cui}}, \bibinfo {author} {\bibfnamefont {W.-J.}\ \bibnamefont {Hu}},
  \bibinfo {author} {\bibfnamefont {X.}~\bibnamefont {Yan}}, \bibinfo {author}
  {\bibfnamefont {C.}~\bibnamefont {Addiego}}, \bibinfo {author} {\bibfnamefont
  {W.}~\bibnamefont {Gao}}, \bibinfo {author} {\bibfnamefont {Y.}~\bibnamefont
  {Wang}}, \bibinfo {author} {\bibfnamefont {Z.}~\bibnamefont {Wang}}, \bibinfo
  {author} {\bibfnamefont {L.}~\bibnamefont {Li}}, \bibinfo {author}
  {\bibfnamefont {Y.}~\bibnamefont {Cheng}}, \bibinfo {author} {\bibfnamefont
  {P.}~\bibnamefont {Li}}, \emph {et~al.},\ }\bibfield  {title} {\bibinfo
  {title} {{Intercorrelated in-plane and out-of-plane ferroelectricity in
  ultrathin two-dimensional layered semiconductor In$_{2}$Se$_{3}$}},\ }\href
  {https://doi.org/10.1021/acs.nanolett.7b04852} {\bibfield  {journal}
  {\bibinfo  {journal} {Nano letters}\ }\textbf {\bibinfo {volume} {18}},\
  \bibinfo {pages} {1253} (\bibinfo {year} {2018})}\BibitemShut {NoStop}%
\bibitem [{\citenamefont {Yuan}\ \emph {et~al.}(2019)\citenamefont {Yuan},
  \citenamefont {Luo}, \citenamefont {Chan}, \citenamefont {Xiao},
  \citenamefont {Dai}, \citenamefont {Xie},\ and\ \citenamefont {Hao}}]{8}%
  \BibitemOpen
  \bibfield  {author} {\bibinfo {author} {\bibfnamefont {S.}~\bibnamefont
  {Yuan}}, \bibinfo {author} {\bibfnamefont {X.}~\bibnamefont {Luo}}, \bibinfo
  {author} {\bibfnamefont {H.~L.}\ \bibnamefont {Chan}}, \bibinfo {author}
  {\bibfnamefont {C.}~\bibnamefont {Xiao}}, \bibinfo {author} {\bibfnamefont
  {Y.}~\bibnamefont {Dai}}, \bibinfo {author} {\bibfnamefont {M.}~\bibnamefont
  {Xie}},\ and\ \bibinfo {author} {\bibfnamefont {J.}~\bibnamefont {Hao}},\
  }\bibfield  {title} {\bibinfo {title} {{Room-temperature ferroelectricity in
  MoTe$_{2}$ down to the atomic monolayer limit}},\ }\href
  {https://doi.org/10.1038/s41467-019-09669-x} {\bibfield  {journal} {\bibinfo
  {journal} {Nature communications}\ }\textbf {\bibinfo {volume} {10}},\
  \bibinfo {pages} {1} (\bibinfo {year} {2019})}\BibitemShut {NoStop}%
\bibitem [{\citenamefont {Song}\ \emph {et~al.}(2022)\citenamefont {Song},
  \citenamefont {Occhialini}, \citenamefont {Erge{\c{c}}en}, \citenamefont
  {Ilyas}, \citenamefont {Amoroso}, \citenamefont {Barone}, \citenamefont
  {Kapeghian}, \citenamefont {Watanabe}, \citenamefont {Taniguchi},
  \citenamefont {Botana} \emph {et~al.}}]{9}%
  \BibitemOpen
  \bibfield  {author} {\bibinfo {author} {\bibfnamefont {Q.}~\bibnamefont
  {Song}}, \bibinfo {author} {\bibfnamefont {C.~A.}\ \bibnamefont
  {Occhialini}}, \bibinfo {author} {\bibfnamefont {E.}~\bibnamefont
  {Erge{\c{c}}en}}, \bibinfo {author} {\bibfnamefont {B.}~\bibnamefont
  {Ilyas}}, \bibinfo {author} {\bibfnamefont {D.}~\bibnamefont {Amoroso}},
  \bibinfo {author} {\bibfnamefont {P.}~\bibnamefont {Barone}}, \bibinfo
  {author} {\bibfnamefont {J.}~\bibnamefont {Kapeghian}}, \bibinfo {author}
  {\bibfnamefont {K.}~\bibnamefont {Watanabe}}, \bibinfo {author}
  {\bibfnamefont {T.}~\bibnamefont {Taniguchi}}, \bibinfo {author}
  {\bibfnamefont {A.~S.}\ \bibnamefont {Botana}}, \emph {et~al.},\ }\bibfield
  {title} {\bibinfo {title} {{Evidence for a single-layer van der Waals
  multiferroic}},\ }\href {https://doi.org/10.1038/s41586-021-04337-x}
  {\bibfield  {journal} {\bibinfo  {journal} {Nature}\ }\textbf {\bibinfo
  {volume} {602}},\ \bibinfo {pages} {601} (\bibinfo {year}
  {2022})}\BibitemShut {NoStop}%
\bibitem [{\citenamefont {Mak}\ \emph {et~al.}(2019)\citenamefont {Mak},
  \citenamefont {Shan},\ and\ \citenamefont {Ralph}}]{10}%
  \BibitemOpen
  \bibfield  {author} {\bibinfo {author} {\bibfnamefont {K.~F.}\ \bibnamefont
  {Mak}}, \bibinfo {author} {\bibfnamefont {J.}~\bibnamefont {Shan}},\ and\
  \bibinfo {author} {\bibfnamefont {D.~C.}\ \bibnamefont {Ralph}},\ }\bibfield
  {title} {\bibinfo {title} {{Probing and controlling magnetic states in 2D
  layered magnetic materials}},\ }\href
  {https://doi.org/10.1038/s42254-019-0110-y} {\bibfield  {journal} {\bibinfo
  {journal} {Nature Reviews Physics}\ }\textbf {\bibinfo {volume} {1}},\
  \bibinfo {pages} {646} (\bibinfo {year} {2019})}\BibitemShut {NoStop}%
\bibitem [{\citenamefont {McGuire}(2017)}]{11}%
  \BibitemOpen
  \bibfield  {author} {\bibinfo {author} {\bibfnamefont {M.~A.}\ \bibnamefont
  {McGuire}},\ }\bibfield  {title} {\bibinfo {title} {Crystal and magnetic
  structures in layered, transition metal dihalides and trihalides},\ }\href
  {https://doi.org/10.3390/cryst7050121} {\bibfield  {journal} {\bibinfo
  {journal} {Crystals}\ }\textbf {\bibinfo {volume} {7}},\ \bibinfo {pages}
  {121} (\bibinfo {year} {2017})}\BibitemShut {NoStop}%
\bibitem [{\citenamefont {Kuindersma}\ \emph {et~al.}(1981)\citenamefont
  {Kuindersma}, \citenamefont {Sanchez},\ and\ \citenamefont {Haas}}]{12}%
  \BibitemOpen
  \bibfield  {author} {\bibinfo {author} {\bibfnamefont {S.}~\bibnamefont
  {Kuindersma}}, \bibinfo {author} {\bibfnamefont {J.}~\bibnamefont
  {Sanchez}},\ and\ \bibinfo {author} {\bibfnamefont {C.}~\bibnamefont
  {Haas}},\ }\bibfield  {title} {\bibinfo {title} {{Magnetic and structural
  investigations on NiI$_{2}$ and CoI$_{2}$}},\ }\href
  {https://doi.org/10.1016/0378-4363(81)90100-5} {\bibfield  {journal}
  {\bibinfo  {journal} {Physica B+ C}\ }\textbf {\bibinfo {volume} {111}},\
  \bibinfo {pages} {231} (\bibinfo {year} {1981})}\BibitemShut {NoStop}%
\bibitem [{\citenamefont {Lu}\ \emph {et~al.}(2019)\citenamefont {Lu},
  \citenamefont {Yao}, \citenamefont {Xiao}, \citenamefont {Huang},\ and\
  \citenamefont {Kan}}]{13}%
  \BibitemOpen
  \bibfield  {author} {\bibinfo {author} {\bibfnamefont {M.}~\bibnamefont
  {Lu}}, \bibinfo {author} {\bibfnamefont {Q.}~\bibnamefont {Yao}}, \bibinfo
  {author} {\bibfnamefont {C.}~\bibnamefont {Xiao}}, \bibinfo {author}
  {\bibfnamefont {C.}~\bibnamefont {Huang}},\ and\ \bibinfo {author}
  {\bibfnamefont {E.}~\bibnamefont {Kan}},\ }\bibfield  {title} {\bibinfo
  {title} {Mechanical, electronic, and magnetic properties of nix$_{2}$ (x= cl,
  br, i) layers},\ }\href {https://doi.org/10.1021/acsomega.9b00056} {\bibfield
   {journal} {\bibinfo  {journal} {ACS omega}\ }\textbf {\bibinfo {volume}
  {4}},\ \bibinfo {pages} {5714} (\bibinfo {year} {2019})}\BibitemShut
  {NoStop}%
\bibitem [{\citenamefont {Liu}\ \emph {et~al.}(2020)\citenamefont {Liu},
  \citenamefont {Wang}, \citenamefont {Wu}, \citenamefont {Chen}, \citenamefont
  {Wan}, \citenamefont {Wen}, \citenamefont {Yang}, \citenamefont {Liu},
  \citenamefont {Song},\ and\ \citenamefont {Xie}}]{14}%
  \BibitemOpen
  \bibfield  {author} {\bibinfo {author} {\bibfnamefont {H.}~\bibnamefont
  {Liu}}, \bibinfo {author} {\bibfnamefont {X.}~\bibnamefont {Wang}}, \bibinfo
  {author} {\bibfnamefont {J.}~\bibnamefont {Wu}}, \bibinfo {author}
  {\bibfnamefont {Y.}~\bibnamefont {Chen}}, \bibinfo {author} {\bibfnamefont
  {J.}~\bibnamefont {Wan}}, \bibinfo {author} {\bibfnamefont {R.}~\bibnamefont
  {Wen}}, \bibinfo {author} {\bibfnamefont {J.}~\bibnamefont {Yang}}, \bibinfo
  {author} {\bibfnamefont {Y.}~\bibnamefont {Liu}}, \bibinfo {author}
  {\bibfnamefont {Z.}~\bibnamefont {Song}},\ and\ \bibinfo {author}
  {\bibfnamefont {L.}~\bibnamefont {Xie}},\ }\bibfield  {title} {\bibinfo
  {title} {{Vapor deposition of magnetic van der Waals NiI$_2$ crystals}},\
  }\href {https://doi.org/10.1021/acsnano.0c04499} {\bibfield  {journal}
  {\bibinfo  {journal} {ACS nano}\ }\textbf {\bibinfo {volume} {14}},\ \bibinfo
  {pages} {10544} (\bibinfo {year} {2020})}\BibitemShut {NoStop}%
\bibitem [{\citenamefont {Amoroso}\ \emph {et~al.}(2021)\citenamefont
  {Amoroso}, \citenamefont {Barone},\ and\ \citenamefont {Picozzi}}]{15}%
  \BibitemOpen
  \bibfield  {author} {\bibinfo {author} {\bibfnamefont {D.}~\bibnamefont
  {Amoroso}}, \bibinfo {author} {\bibfnamefont {P.}~\bibnamefont {Barone}},\
  and\ \bibinfo {author} {\bibfnamefont {S.}~\bibnamefont {Picozzi}},\
  }\bibfield  {title} {\bibinfo {title} {{Interplay between Single-Ion and
  Two-Ion Anisotropies in Frustrated 2D Semiconductors and Tuning of Magnetic
  Structures Topology}},\ }\href {https://doi.org/10.3390/nano11081873}
  {\bibfield  {journal} {\bibinfo  {journal} {Nanomaterials}\ }\textbf
  {\bibinfo {volume} {11}},\ \bibinfo {pages} {1873} (\bibinfo {year}
  {2021})}\BibitemShut {NoStop}%
\bibitem [{\citenamefont {Kurumaji}\ \emph
  {et~al.}(2013{\natexlab{a}})\citenamefont {Kurumaji}, \citenamefont {Seki},
  \citenamefont {Ishiwata}, \citenamefont {Murakawa}, \citenamefont {Kaneko},\
  and\ \citenamefont {Tokura}}]{16}%
  \BibitemOpen
  \bibfield  {author} {\bibinfo {author} {\bibfnamefont {T.}~\bibnamefont
  {Kurumaji}}, \bibinfo {author} {\bibfnamefont {S.}~\bibnamefont {Seki}},
  \bibinfo {author} {\bibfnamefont {S.}~\bibnamefont {Ishiwata}}, \bibinfo
  {author} {\bibfnamefont {H.}~\bibnamefont {Murakawa}}, \bibinfo {author}
  {\bibfnamefont {Y.}~\bibnamefont {Kaneko}},\ and\ \bibinfo {author}
  {\bibfnamefont {Y.}~\bibnamefont {Tokura}},\ }\bibfield  {title} {\bibinfo
  {title} {{Magnetoelectric responses induced by domain rearrangement and spin
  structural change in triangular-lattice helimagnets NiI${}_{2}$ and
  CoI${}_{2}$}},\ }\href {https://doi.org/10.1103/PhysRevB.87.014429}
  {\bibfield  {journal} {\bibinfo  {journal} {Phys. Rev. B}\ }\textbf {\bibinfo
  {volume} {87}},\ \bibinfo {pages} {014429} (\bibinfo {year}
  {2013}{\natexlab{a}})}\BibitemShut {NoStop}%
\bibitem [{\citenamefont {Ju}\ \emph {et~al.}(2021{\natexlab{a}})\citenamefont
  {Ju}, \citenamefont {Lee}, \citenamefont {Kim}, \citenamefont {Choi},
  \citenamefont {Roh}, \citenamefont {Son}, \citenamefont {Park}, \citenamefont
  {Kim}, \citenamefont {Jung}, \citenamefont {Kim} \emph {et~al.}}]{17}%
  \BibitemOpen
  \bibfield  {author} {\bibinfo {author} {\bibfnamefont {H.}~\bibnamefont
  {Ju}}, \bibinfo {author} {\bibfnamefont {Y.}~\bibnamefont {Lee}}, \bibinfo
  {author} {\bibfnamefont {K.-T.}\ \bibnamefont {Kim}}, \bibinfo {author}
  {\bibfnamefont {I.~H.}\ \bibnamefont {Choi}}, \bibinfo {author}
  {\bibfnamefont {C.~J.}\ \bibnamefont {Roh}}, \bibinfo {author} {\bibfnamefont
  {S.}~\bibnamefont {Son}}, \bibinfo {author} {\bibfnamefont {P.}~\bibnamefont
  {Park}}, \bibinfo {author} {\bibfnamefont {J.~H.}\ \bibnamefont {Kim}},
  \bibinfo {author} {\bibfnamefont {T.~S.}\ \bibnamefont {Jung}}, \bibinfo
  {author} {\bibfnamefont {J.~H.}\ \bibnamefont {Kim}}, \emph {et~al.},\
  }\bibfield  {title} {\bibinfo {title} {{Possible persistence of multiferroic
  order down to bilayer limit of van der Waals material NiI$_2$}},\ }\href
  {https://doi.org/10.1021/acs.nanolett.1c01095} {\bibfield  {journal}
  {\bibinfo  {journal} {Nano Letters}\ }\textbf {\bibinfo {volume} {21}},\
  \bibinfo {pages} {5126} (\bibinfo {year} {2021}{\natexlab{a}})}\BibitemShut
  {NoStop}%
\bibitem [{\citenamefont {Li}\ \emph {et~al.}(2019)\citenamefont {Li},
  \citenamefont {Jiang}, \citenamefont {Sivadas}, \citenamefont {Wang},
  \citenamefont {Xu}, \citenamefont {Weber}, \citenamefont {Goldberger},
  \citenamefont {Watanabe}, \citenamefont {Taniguchi}, \citenamefont {Fennie}
  \emph {et~al.}}]{20}%
  \BibitemOpen
  \bibfield  {author} {\bibinfo {author} {\bibfnamefont {T.}~\bibnamefont
  {Li}}, \bibinfo {author} {\bibfnamefont {S.}~\bibnamefont {Jiang}}, \bibinfo
  {author} {\bibfnamefont {N.}~\bibnamefont {Sivadas}}, \bibinfo {author}
  {\bibfnamefont {Z.}~\bibnamefont {Wang}}, \bibinfo {author} {\bibfnamefont
  {Y.}~\bibnamefont {Xu}}, \bibinfo {author} {\bibfnamefont {D.}~\bibnamefont
  {Weber}}, \bibinfo {author} {\bibfnamefont {J.~E.}\ \bibnamefont
  {Goldberger}}, \bibinfo {author} {\bibfnamefont {K.}~\bibnamefont
  {Watanabe}}, \bibinfo {author} {\bibfnamefont {T.}~\bibnamefont {Taniguchi}},
  \bibinfo {author} {\bibfnamefont {C.~J.}\ \bibnamefont {Fennie}}, \emph
  {et~al.},\ }\bibfield  {title} {\bibinfo {title} {{Pressure-controlled
  interlayer magnetism in atomically thin CrI$_3$}},\ }\href
  {https://doi.org/10.1038/s41563-019-0506-1} {\bibfield  {journal} {\bibinfo
  {journal} {Nature materials}\ }\textbf {\bibinfo {volume} {18}},\ \bibinfo
  {pages} {1303} (\bibinfo {year} {2019})}\BibitemShut {NoStop}%
\bibitem [{\citenamefont {Wang}\ \emph {et~al.}(2020)\citenamefont {Wang},
  \citenamefont {Wang}, \citenamefont {Liang}, \citenamefont {Ma},
  \citenamefont {Xu}, \citenamefont {Liu}, \citenamefont {Zhang}, \citenamefont
  {Admasu}, \citenamefont {Cheong}, \citenamefont {Wang} \emph {et~al.}}]{21}%
  \BibitemOpen
  \bibfield  {author} {\bibinfo {author} {\bibfnamefont {Y.}~\bibnamefont
  {Wang}}, \bibinfo {author} {\bibfnamefont {C.}~\bibnamefont {Wang}}, \bibinfo
  {author} {\bibfnamefont {S.-J.}\ \bibnamefont {Liang}}, \bibinfo {author}
  {\bibfnamefont {Z.}~\bibnamefont {Ma}}, \bibinfo {author} {\bibfnamefont
  {K.}~\bibnamefont {Xu}}, \bibinfo {author} {\bibfnamefont {X.}~\bibnamefont
  {Liu}}, \bibinfo {author} {\bibfnamefont {L.}~\bibnamefont {Zhang}}, \bibinfo
  {author} {\bibfnamefont {A.~S.}\ \bibnamefont {Admasu}}, \bibinfo {author}
  {\bibfnamefont {S.-W.}\ \bibnamefont {Cheong}}, \bibinfo {author}
  {\bibfnamefont {L.}~\bibnamefont {Wang}}, \emph {et~al.},\ }\bibfield
  {title} {\bibinfo {title} {Strain-sensitive magnetization reversal of a van
  der waals magnet},\ }\href {https://doi.org/10.1002/adma.202004533}
  {\bibfield  {journal} {\bibinfo  {journal} {Advanced Materials}\ }\textbf
  {\bibinfo {volume} {32}},\ \bibinfo {pages} {2004533} (\bibinfo {year}
  {2020})}\BibitemShut {NoStop}%
\bibitem [{\citenamefont {Wu}\ \emph {et~al.}(2022)\citenamefont {Wu},
  \citenamefont {Zhou}, \citenamefont {Zhou}, \citenamefont {Wang},\ and\
  \citenamefont {Ji}}]{22}%
  \BibitemOpen
  \bibfield  {author} {\bibinfo {author} {\bibfnamefont {L.}~\bibnamefont
  {Wu}}, \bibinfo {author} {\bibfnamefont {L.}~\bibnamefont {Zhou}}, \bibinfo
  {author} {\bibfnamefont {X.}~\bibnamefont {Zhou}}, \bibinfo {author}
  {\bibfnamefont {C.}~\bibnamefont {Wang}},\ and\ \bibinfo {author}
  {\bibfnamefont {W.}~\bibnamefont {Ji}},\ }\bibfield  {title} {\bibinfo
  {title} {In-plane epitaxy-strain-tuning intralayer and interlayer magnetic
  coupling in ${\mathrm{crse}}_{2}$ and ${\mathrm{crte}}_{2}$ monolayers and
  bilayers},\ }\href {https://doi.org/10.1103/PhysRevB.106.L081401} {\bibfield
  {journal} {\bibinfo  {journal} {Phys. Rev. B}\ }\textbf {\bibinfo {volume}
  {106}},\ \bibinfo {pages} {L081401} (\bibinfo {year} {2022})}\BibitemShut
  {NoStop}%
\bibitem [{\citenamefont {Kresse}\ and\ \citenamefont {Hafner}(1993)}]{23}%
  \BibitemOpen
  \bibfield  {author} {\bibinfo {author} {\bibfnamefont {G.}~\bibnamefont
  {Kresse}}\ and\ \bibinfo {author} {\bibfnamefont {J.}~\bibnamefont
  {Hafner}},\ }\bibfield  {title} {\bibinfo {title} {Ab initio molecular
  dynamics for liquid metals},\ }\href
  {https://doi.org/10.1103/PhysRevB.47.558} {\bibfield  {journal} {\bibinfo
  {journal} {Phys. Rev. B}\ }\textbf {\bibinfo {volume} {47}},\ \bibinfo
  {pages} {558} (\bibinfo {year} {1993})}\BibitemShut {NoStop}%
\bibitem [{\citenamefont {Kresse}\ and\ \citenamefont
  {Furthm\"uller}(1996)}]{24}%
  \BibitemOpen
  \bibfield  {author} {\bibinfo {author} {\bibfnamefont {G.}~\bibnamefont
  {Kresse}}\ and\ \bibinfo {author} {\bibfnamefont {J.}~\bibnamefont
  {Furthm\"uller}},\ }\bibfield  {title} {\bibinfo {title} {Efficient iterative
  schemes for ab initio total-energy calculations using a plane-wave basis
  set},\ }\href {https://doi.org/10.1103/PhysRevB.54.11169} {\bibfield
  {journal} {\bibinfo  {journal} {Phys. Rev. B}\ }\textbf {\bibinfo {volume}
  {54}},\ \bibinfo {pages} {11169} (\bibinfo {year} {1996})}\BibitemShut
  {NoStop}%
\bibitem [{\citenamefont {Bl\"ochl}(1994)}]{26}%
  \BibitemOpen
  \bibfield  {author} {\bibinfo {author} {\bibfnamefont {P.~E.}\ \bibnamefont
  {Bl\"ochl}},\ }\bibfield  {title} {\bibinfo {title} {Projector augmented-wave
  method},\ }\href {https://doi.org/10.1103/PhysRevB.50.17953} {\bibfield
  {journal} {\bibinfo  {journal} {Phys. Rev. B}\ }\textbf {\bibinfo {volume}
  {50}},\ \bibinfo {pages} {17953} (\bibinfo {year} {1994})}\BibitemShut
  {NoStop}%
\bibitem [{\citenamefont {Liechtenstein}\ \emph {et~al.}(1995)\citenamefont
  {Liechtenstein}, \citenamefont {Anisimov},\ and\ \citenamefont
  {Zaanen}}]{25}%
  \BibitemOpen
  \bibfield  {author} {\bibinfo {author} {\bibfnamefont {A.~I.}\ \bibnamefont
  {Liechtenstein}}, \bibinfo {author} {\bibfnamefont {V.~I.}\ \bibnamefont
  {Anisimov}},\ and\ \bibinfo {author} {\bibfnamefont {J.}~\bibnamefont
  {Zaanen}},\ }\bibfield  {title} {\bibinfo {title} {{Density-functional theory
  and strong interactions: Orbital ordering in Mott-Hubbard insulators}},\
  }\href {https://doi.org/10.1103/PhysRevB.52.R5467} {\bibfield  {journal}
  {\bibinfo  {journal} {Phys. Rev. B}\ }\textbf {\bibinfo {volume} {52}},\
  \bibinfo {pages} {R5467} (\bibinfo {year} {1995})}\BibitemShut {NoStop}%
\bibitem [{\citenamefont {Chen}\ \emph {et~al.}(1993)\citenamefont {Chen},
  \citenamefont {Yu},\ and\ \citenamefont {Taylor}}]{28}%
  \BibitemOpen
  \bibfield  {author} {\bibinfo {author} {\bibfnamefont {A.~L.}\ \bibnamefont
  {Chen}}, \bibinfo {author} {\bibfnamefont {P.~Y.}\ \bibnamefont {Yu}},\ and\
  \bibinfo {author} {\bibfnamefont {R.~D.}\ \bibnamefont {Taylor}},\ }\bibfield
   {title} {\bibinfo {title} {{Closure of the charge-transfer energy gap and
  metallization of NiI$_{2}$ under pressure}},\ }\href
  {https://doi.org/10.1103/PhysRevLett.71.4011} {\bibfield  {journal} {\bibinfo
   {journal} {Phys. Rev. Lett.}\ }\textbf {\bibinfo {volume} {71}},\ \bibinfo
  {pages} {4011} (\bibinfo {year} {1993})}\BibitemShut {NoStop}%
\bibitem [{\citenamefont {Cui}\ \emph {et~al.}(2020)\citenamefont {Cui},
  \citenamefont {Liang}, \citenamefont {Yang}, \citenamefont {Wang},
  \citenamefont {Li}, \citenamefont {Cui},\ and\ \citenamefont {Yang}}]{29}%
  \BibitemOpen
  \bibfield  {author} {\bibinfo {author} {\bibfnamefont {Q.}~\bibnamefont
  {Cui}}, \bibinfo {author} {\bibfnamefont {J.}~\bibnamefont {Liang}}, \bibinfo
  {author} {\bibfnamefont {B.}~\bibnamefont {Yang}}, \bibinfo {author}
  {\bibfnamefont {Z.}~\bibnamefont {Wang}}, \bibinfo {author} {\bibfnamefont
  {P.}~\bibnamefont {Li}}, \bibinfo {author} {\bibfnamefont {P.}~\bibnamefont
  {Cui}},\ and\ \bibinfo {author} {\bibfnamefont {H.}~\bibnamefont {Yang}},\
  }\bibfield  {title} {\bibinfo {title} {{Giant enhancement of perpendicular
  magnetic anisotropy and induced quantum anomalous Hall effect in
  graphene/NiI$_{2}$ heterostructures via tuning the van der Waals interlayer
  distance}},\ }\href {https://doi.org/10.1103/PhysRevB.101.214439} {\bibfield
  {journal} {\bibinfo  {journal} {Phys. Rev. B}\ }\textbf {\bibinfo {volume}
  {101}},\ \bibinfo {pages} {214439} (\bibinfo {year} {2020})}\BibitemShut
  {NoStop}%
\bibitem [{\citenamefont {Morgan}\ \emph {et~al.}(2003)\citenamefont {Morgan},
  \citenamefont {Wang}, \citenamefont {Ceder},\ and\ \citenamefont {van~de
  Walle}}]{30}%
  \BibitemOpen
  \bibfield  {author} {\bibinfo {author} {\bibfnamefont {D.}~\bibnamefont
  {Morgan}}, \bibinfo {author} {\bibfnamefont {B.}~\bibnamefont {Wang}},
  \bibinfo {author} {\bibfnamefont {G.}~\bibnamefont {Ceder}},\ and\ \bibinfo
  {author} {\bibfnamefont {A.}~\bibnamefont {van~de Walle}},\ }\bibfield
  {title} {\bibinfo {title} {{First-principles study of magnetism in spinel
  MnO$_{2}$}},\ }\href {https://doi.org/10.1103/PhysRevB.67.134404} {\bibfield
  {journal} {\bibinfo  {journal} {Phys. Rev. B}\ }\textbf {\bibinfo {volume}
  {67}},\ \bibinfo {pages} {134404} (\bibinfo {year} {2003})}\BibitemShut
  {NoStop}%
\bibitem [{\citenamefont {Fedorova}\ \emph {et~al.}(2015)\citenamefont
  {Fedorova}, \citenamefont {Ederer}, \citenamefont {Spaldin},\ and\
  \citenamefont {Scaramucci}}]{31}%
  \BibitemOpen
  \bibfield  {author} {\bibinfo {author} {\bibfnamefont {N.~S.}\ \bibnamefont
  {Fedorova}}, \bibinfo {author} {\bibfnamefont {C.}~\bibnamefont {Ederer}},
  \bibinfo {author} {\bibfnamefont {N.~A.}\ \bibnamefont {Spaldin}},\ and\
  \bibinfo {author} {\bibfnamefont {A.}~\bibnamefont {Scaramucci}},\ }\bibfield
   {title} {\bibinfo {title} {Biquadratic and ring exchange interactions in
  orthorhombic perovskite manganites},\ }\href
  {https://doi.org/10.1103/PhysRevB.91.165122} {\bibfield  {journal} {\bibinfo
  {journal} {Phys. Rev. B}\ }\textbf {\bibinfo {volume} {91}},\ \bibinfo
  {pages} {165122} (\bibinfo {year} {2015})}\BibitemShut {NoStop}%
\bibitem [{\citenamefont {Liu}\ \emph {et~al.}(2022)\citenamefont {Liu},
  \citenamefont {Ni}, \citenamefont {Li}, \citenamefont {Sun}, \citenamefont
  {Liang}, \citenamefont {Frandsen}, \citenamefont {Christianson},
  \citenamefont {dela Cruz}, \citenamefont {Xu}, \citenamefont {Yao},
  \citenamefont {Lynn}, \citenamefont {Birgeneau}, \citenamefont {Cao},\ and\
  \citenamefont {Wang}}]{PhysRevB.105.214303}%
  \BibitemOpen
  \bibfield  {author} {\bibinfo {author} {\bibfnamefont {Z.}~\bibnamefont
  {Liu}}, \bibinfo {author} {\bibfnamefont {X.-S.}\ \bibnamefont {Ni}},
  \bibinfo {author} {\bibfnamefont {L.}~\bibnamefont {Li}}, \bibinfo {author}
  {\bibfnamefont {H.}~\bibnamefont {Sun}}, \bibinfo {author} {\bibfnamefont
  {F.}~\bibnamefont {Liang}}, \bibinfo {author} {\bibfnamefont {B.~A.}\
  \bibnamefont {Frandsen}}, \bibinfo {author} {\bibfnamefont {A.~D.}\
  \bibnamefont {Christianson}}, \bibinfo {author} {\bibfnamefont
  {C.}~\bibnamefont {dela Cruz}}, \bibinfo {author} {\bibfnamefont
  {Z.}~\bibnamefont {Xu}}, \bibinfo {author} {\bibfnamefont {D.-X.}\
  \bibnamefont {Yao}}, \bibinfo {author} {\bibfnamefont {J.~W.}\ \bibnamefont
  {Lynn}}, \bibinfo {author} {\bibfnamefont {R.~J.}\ \bibnamefont {Birgeneau}},
  \bibinfo {author} {\bibfnamefont {K.}~\bibnamefont {Cao}},\ and\ \bibinfo
  {author} {\bibfnamefont {M.}~\bibnamefont {Wang}},\ }\bibfield  {title}
  {\bibinfo {title} {Effect of iron vacancies on magnetic order and spin
  dynamics of the spin ladder
  ${\mathrm{bafe}}_{2\ensuremath{-}\ensuremath{\delta}}{\mathrm{s}}_{1.5}{\mathrm{se}}_{1.5}$},\
  }\href {https://doi.org/10.1103/PhysRevB.105.214303} {\bibfield  {journal}
  {\bibinfo  {journal} {Phys. Rev. B}\ }\textbf {\bibinfo {volume} {105}},\
  \bibinfo {pages} {214303} (\bibinfo {year} {2022})}\BibitemShut {NoStop}%
\bibitem [{\citenamefont {King-Smith}\ and\ \citenamefont
  {Vanderbilt}(1993)}]{27}%
  \BibitemOpen
  \bibfield  {author} {\bibinfo {author} {\bibfnamefont {R.~D.}\ \bibnamefont
  {King-Smith}}\ and\ \bibinfo {author} {\bibfnamefont {D.}~\bibnamefont
  {Vanderbilt}},\ }\bibfield  {title} {\bibinfo {title} {Theory of polarization
  of crystalline solids},\ }\href {https://doi.org/10.1103/PhysRevB.47.1651}
  {\bibfield  {journal} {\bibinfo  {journal} {Phys. Rev. B}\ }\textbf {\bibinfo
  {volume} {47}},\ \bibinfo {pages} {1651} (\bibinfo {year}
  {1993})}\BibitemShut {NoStop}%
\bibitem [{\citenamefont {Grimme}\ \emph {et~al.}(2010)\citenamefont {Grimme},
  \citenamefont {Antony}, \citenamefont {Ehrlich},\ and\ \citenamefont
  {Krieg}}]{grimme2010consistent}%
  \BibitemOpen
  \bibfield  {author} {\bibinfo {author} {\bibfnamefont {S.}~\bibnamefont
  {Grimme}}, \bibinfo {author} {\bibfnamefont {J.}~\bibnamefont {Antony}},
  \bibinfo {author} {\bibfnamefont {S.}~\bibnamefont {Ehrlich}},\ and\ \bibinfo
  {author} {\bibfnamefont {H.}~\bibnamefont {Krieg}},\ }\bibfield  {title}
  {\bibinfo {title} {{A consistent and accurate ab initio parametrization of
  density functional dispersion correction (DFT-D) for the 94 elements H-Pu}},\
  }\href {https://doi.org/10.1063/1.3382344} {\bibfield  {journal} {\bibinfo
  {journal} {The Journal of chemical physics}\ }\textbf {\bibinfo {volume}
  {132}},\ \bibinfo {pages} {154104} (\bibinfo {year} {2010})}\BibitemShut
  {NoStop}%
\bibitem [{\citenamefont {Cao}\ \emph {et~al.}(2009)\citenamefont {Cao},
  \citenamefont {Guo}, \citenamefont {Vanderbilt},\ and\ \citenamefont
  {He}}]{32}%
  \BibitemOpen
  \bibfield  {author} {\bibinfo {author} {\bibfnamefont {K.}~\bibnamefont
  {Cao}}, \bibinfo {author} {\bibfnamefont {G.-C.}\ \bibnamefont {Guo}},
  \bibinfo {author} {\bibfnamefont {D.}~\bibnamefont {Vanderbilt}},\ and\
  \bibinfo {author} {\bibfnamefont {L.}~\bibnamefont {He}},\ }\bibfield
  {title} {\bibinfo {title} {First-principles modeling of multiferroic
  $r{\mathrm{mn}}_{2}{\mathbf{o}}_{5}$},\ }\href
  {https://doi.org/10.1103/PhysRevLett.103.257201} {\bibfield  {journal}
  {\bibinfo  {journal} {Phys. Rev. Lett.}\ }\textbf {\bibinfo {volume} {103}},\
  \bibinfo {pages} {257201} (\bibinfo {year} {2009})}\BibitemShut {NoStop}%
\bibitem [{\citenamefont {Ju}\ \emph {et~al.}(2021{\natexlab{b}})\citenamefont
  {Ju}, \citenamefont {Lee}, \citenamefont {Kim}, \citenamefont {Choi},
  \citenamefont {Roh}, \citenamefont {Son}, \citenamefont {Park}, \citenamefont
  {Kim}, \citenamefont {Jung}, \citenamefont {Kim} \emph {et~al.}}]{32-1}%
  \BibitemOpen
  \bibfield  {author} {\bibinfo {author} {\bibfnamefont {H.}~\bibnamefont
  {Ju}}, \bibinfo {author} {\bibfnamefont {Y.}~\bibnamefont {Lee}}, \bibinfo
  {author} {\bibfnamefont {K.-T.}\ \bibnamefont {Kim}}, \bibinfo {author}
  {\bibfnamefont {I.~H.}\ \bibnamefont {Choi}}, \bibinfo {author}
  {\bibfnamefont {C.~J.}\ \bibnamefont {Roh}}, \bibinfo {author} {\bibfnamefont
  {S.}~\bibnamefont {Son}}, \bibinfo {author} {\bibfnamefont {P.}~\bibnamefont
  {Park}}, \bibinfo {author} {\bibfnamefont {J.~H.}\ \bibnamefont {Kim}},
  \bibinfo {author} {\bibfnamefont {T.~S.}\ \bibnamefont {Jung}}, \bibinfo
  {author} {\bibfnamefont {J.~H.}\ \bibnamefont {Kim}}, \emph {et~al.},\
  }\bibfield  {title} {\bibinfo {title} {{Possible persistence of multiferroic
  order down to bilayer limit of van der Waals material NiI$_2$}},\ }\href
  {https://doi.org/10.1021/acs.nanolett.1c01095} {\bibfield  {journal}
  {\bibinfo  {journal} {Nano Letters}\ }\textbf {\bibinfo {volume} {21}},\
  \bibinfo {pages} {5126} (\bibinfo {year} {2021}{\natexlab{b}})}\BibitemShut
  {NoStop}%
\bibitem [{\citenamefont {An}\ \emph {et~al.}(2022{\natexlab{a}})\citenamefont
  {An}, \citenamefont {Wang}, \citenamefont {Liao}, \citenamefont {Gao},
  \citenamefont {Chen}, \citenamefont {Wu}, \citenamefont {Li}, \citenamefont
  {Xu},\ and\ \citenamefont {Ma}}]{32-2}%
  \BibitemOpen
  \bibfield  {author} {\bibinfo {author} {\bibfnamefont {Y.}~\bibnamefont
  {An}}, \bibinfo {author} {\bibfnamefont {H.}~\bibnamefont {Wang}}, \bibinfo
  {author} {\bibfnamefont {J.}~\bibnamefont {Liao}}, \bibinfo {author}
  {\bibfnamefont {Y.}~\bibnamefont {Gao}}, \bibinfo {author} {\bibfnamefont
  {J.}~\bibnamefont {Chen}}, \bibinfo {author} {\bibfnamefont {Y.}~\bibnamefont
  {Wu}}, \bibinfo {author} {\bibfnamefont {Y.}~\bibnamefont {Li}}, \bibinfo
  {author} {\bibfnamefont {G.}~\bibnamefont {Xu}},\ and\ \bibinfo {author}
  {\bibfnamefont {C.}~\bibnamefont {Ma}},\ }\bibfield  {title} {\bibinfo
  {title} {{Spin transport properties and nanodevice simulations of NiI$_2$
  monolayer}},\ }\href {https://doi.org/10.1016/j.physe.2022.115262} {\bibfield
   {journal} {\bibinfo  {journal} {Physica E: Low-dimensional Systems and
  Nanostructures}\ }\textbf {\bibinfo {volume} {142}},\ \bibinfo {pages}
  {115262} (\bibinfo {year} {2022}{\natexlab{a}})}\BibitemShut {NoStop}%
\bibitem [{\citenamefont {Goodenough}(2008)}]{33}%
  \BibitemOpen
  \bibfield  {author} {\bibinfo {author} {\bibfnamefont {J.~B.}\ \bibnamefont
  {Goodenough}},\ }\bibfield  {title} {\bibinfo {title} {{Goodenough-Kanamori
  rule}},\ }\href {https://doi.org/10.4249/scholarpedia.7382} {\bibfield
  {journal} {\bibinfo  {journal} {Scholarpedia}\ }\textbf {\bibinfo {volume}
  {3}},\ \bibinfo {pages} {7382} (\bibinfo {year} {2008})}\BibitemShut
  {NoStop}%
\bibitem [{\citenamefont {Kurumaji}\ \emph
  {et~al.}(2013{\natexlab{b}})\citenamefont {Kurumaji}, \citenamefont {Seki},
  \citenamefont {Ishiwata}, \citenamefont {Murakawa}, \citenamefont {Kaneko},\
  and\ \citenamefont {Tokura}}]{32-3}%
  \BibitemOpen
  \bibfield  {author} {\bibinfo {author} {\bibfnamefont {T.}~\bibnamefont
  {Kurumaji}}, \bibinfo {author} {\bibfnamefont {S.}~\bibnamefont {Seki}},
  \bibinfo {author} {\bibfnamefont {S.}~\bibnamefont {Ishiwata}}, \bibinfo
  {author} {\bibfnamefont {H.}~\bibnamefont {Murakawa}}, \bibinfo {author}
  {\bibfnamefont {Y.}~\bibnamefont {Kaneko}},\ and\ \bibinfo {author}
  {\bibfnamefont {Y.}~\bibnamefont {Tokura}},\ }\bibfield  {title} {\bibinfo
  {title} {{Magnetoelectric responses induced by domain rearrangement and spin
  structural change in triangular-lattice helimagnets NiI$_2$ and CoI$_2$}},\
  }\href {https://doi.org/10.1103/PhysRevB.87.014429} {\bibfield  {journal}
  {\bibinfo  {journal} {Physical Review B}\ }\textbf {\bibinfo {volume} {87}},\
  \bibinfo {pages} {014429} (\bibinfo {year} {2013}{\natexlab{b}})}\BibitemShut
  {NoStop}%
\bibitem [{\citenamefont {Fumega}\ and\ \citenamefont {Lado}(2022)}]{19}%
  \BibitemOpen
  \bibfield  {author} {\bibinfo {author} {\bibfnamefont {A.~O.}\ \bibnamefont
  {Fumega}}\ and\ \bibinfo {author} {\bibfnamefont {J.~L.}\ \bibnamefont
  {Lado}},\ }\bibfield  {title} {\bibinfo {title} {Microscopic origin of
  multiferroic order in monolayer {NiI}$_2$},\ }\href
  {https://doi.org/10.1088/2053-1583/ac4e9d} {\bibfield  {journal} {\bibinfo
  {journal} {2D Materials}\ }\textbf {\bibinfo {volume} {9}},\ \bibinfo {pages}
  {025010} (\bibinfo {year} {2022})}\BibitemShut {NoStop}%
\bibitem [{\citenamefont {Amoroso}\ \emph {et~al.}(2020)\citenamefont
  {Amoroso}, \citenamefont {Barone},\ and\ \citenamefont {Picozzi}}]{34}%
  \BibitemOpen
  \bibfield  {author} {\bibinfo {author} {\bibfnamefont {D.}~\bibnamefont
  {Amoroso}}, \bibinfo {author} {\bibfnamefont {P.}~\bibnamefont {Barone}},\
  and\ \bibinfo {author} {\bibfnamefont {S.}~\bibnamefont {Picozzi}},\
  }\bibfield  {title} {\bibinfo {title} {{Spontaneous skyrmionic lattice from
  anisotropic symmetric exchange in a Ni-halide monolayer}},\ }\href
  {https://doi.org/10.1038/s41467-020-19535-w} {\bibfield  {journal} {\bibinfo
  {journal} {Nature communications}\ }\textbf {\bibinfo {volume} {11}},\
  \bibinfo {pages} {1} (\bibinfo {year} {2020})}\BibitemShut {NoStop}%
\bibitem [{\citenamefont {Xiang}\ \emph {et~al.}(2011)\citenamefont {Xiang},
  \citenamefont {Kan}, \citenamefont {Zhang}, \citenamefont {Whangbo},\ and\
  \citenamefont {Gong}}]{18}%
  \BibitemOpen
  \bibfield  {author} {\bibinfo {author} {\bibfnamefont {H.~J.}\ \bibnamefont
  {Xiang}}, \bibinfo {author} {\bibfnamefont {E.~J.}\ \bibnamefont {Kan}},
  \bibinfo {author} {\bibfnamefont {Y.}~\bibnamefont {Zhang}}, \bibinfo
  {author} {\bibfnamefont {M.-H.}\ \bibnamefont {Whangbo}},\ and\ \bibinfo
  {author} {\bibfnamefont {X.~G.}\ \bibnamefont {Gong}},\ }\bibfield  {title}
  {\bibinfo {title} {General theory for the ferroelectric polarization induced
  by spin-spiral order},\ }\href
  {https://doi.org/10.1103/PhysRevLett.107.157202} {\bibfield  {journal}
  {\bibinfo  {journal} {Phys. Rev. Lett.}\ }\textbf {\bibinfo {volume} {107}},\
  \bibinfo {pages} {157202} (\bibinfo {year} {2011})}\BibitemShut {NoStop}%
\bibitem [{\citenamefont {An}\ \emph {et~al.}(2022{\natexlab{b}})\citenamefont
  {An}, \citenamefont {Wang}, \citenamefont {Liao}, \citenamefont {Gao},
  \citenamefont {Chen}, \citenamefont {Wu}, \citenamefont {Li}, \citenamefont
  {Xu},\ and\ \citenamefont {Ma}}]{AN2022115262}%
  \BibitemOpen
  \bibfield  {author} {\bibinfo {author} {\bibfnamefont {Y.}~\bibnamefont
  {An}}, \bibinfo {author} {\bibfnamefont {H.}~\bibnamefont {Wang}}, \bibinfo
  {author} {\bibfnamefont {J.}~\bibnamefont {Liao}}, \bibinfo {author}
  {\bibfnamefont {Y.}~\bibnamefont {Gao}}, \bibinfo {author} {\bibfnamefont
  {J.}~\bibnamefont {Chen}}, \bibinfo {author} {\bibfnamefont {Y.}~\bibnamefont
  {Wu}}, \bibinfo {author} {\bibfnamefont {Y.}~\bibnamefont {Li}}, \bibinfo
  {author} {\bibfnamefont {G.}~\bibnamefont {Xu}},\ and\ \bibinfo {author}
  {\bibfnamefont {C.}~\bibnamefont {Ma}},\ }\bibfield  {title} {\bibinfo
  {title} {Spin transport properties and nanodevice simulations of nii2
  monolayer},\ }\href
  {https://doi.org/https://doi.org/10.1016/j.physe.2022.115262} {\bibfield
  {journal} {\bibinfo  {journal} {Physica E: Low-dimensional Systems and
  Nanostructures}\ }\textbf {\bibinfo {volume} {142}},\ \bibinfo {pages}
  {115262} (\bibinfo {year} {2022}{\natexlab{b}})}\BibitemShut {NoStop}%
\bibitem [{\citenamefont {Dong}\ \emph {et~al.}(2022)\citenamefont {Dong},
  \citenamefont {Ren},\ and\ \citenamefont {Zhang}}]{dong2022quantum}%
  \BibitemOpen
  \bibfield  {author} {\bibinfo {author} {\bibfnamefont {X.-j.}\ \bibnamefont
  {Dong}}, \bibinfo {author} {\bibfnamefont {M.-j.}\ \bibnamefont {Ren}},\ and\
  \bibinfo {author} {\bibfnamefont {C.-w.}\ \bibnamefont {Zhang}},\ }\bibfield
  {title} {\bibinfo {title} {Quantum anomalous hall effect in germanene by
  proximity coupling to a semiconducting ferromagnetic substrate nii2},\
  }\bibfield  {journal} {\bibinfo  {journal} {Physical Chemistry Chemical
  Physics}\ }\href {https://doi.org/10.1039/D2CP02688K} {10.1039/D2CP02688K}
  (\bibinfo {year} {2022})\BibitemShut {NoStop}%
\bibitem [{\citenamefont {Boudjelal}\ \emph {et~al.}(2019)\citenamefont
  {Boudjelal}, \citenamefont {Belfedal}, \citenamefont {Bouadjemi},
  \citenamefont {Lantri}, \citenamefont {Bentata}, \citenamefont {Batouche},\
  and\ \citenamefont {Khenata}}]{35}%
  \BibitemOpen
  \bibfield  {author} {\bibinfo {author} {\bibfnamefont {M.}~\bibnamefont
  {Boudjelal}}, \bibinfo {author} {\bibfnamefont {A.}~\bibnamefont {Belfedal}},
  \bibinfo {author} {\bibfnamefont {B.}~\bibnamefont {Bouadjemi}}, \bibinfo
  {author} {\bibfnamefont {T.}~\bibnamefont {Lantri}}, \bibinfo {author}
  {\bibfnamefont {R.}~\bibnamefont {Bentata}}, \bibinfo {author} {\bibfnamefont
  {M.}~\bibnamefont {Batouche}},\ and\ \bibinfo {author} {\bibfnamefont
  {R.}~\bibnamefont {Khenata}},\ }\bibfield  {title} {\bibinfo {title}
  {{Ferromagnetic Half-Semiconductor (HSC) gaps in co-doped CdS: Ab-initio
  study}},\ }\href {https://doi.org/10.1016/j.cjph.2019.09.004} {\bibfield
  {journal} {\bibinfo  {journal} {Chinese Journal of Physics}\ }\textbf
  {\bibinfo {volume} {61}},\ \bibinfo {pages} {155} (\bibinfo {year}
  {2019})}\BibitemShut {NoStop}%
\bibitem [{\citenamefont {Ku}\ \emph {et~al.}(2010)\citenamefont {Ku},
  \citenamefont {Berlijn},\ and\ \citenamefont {Lee}}]{PhysRevLett.104.216401}%
  \BibitemOpen
  \bibfield  {author} {\bibinfo {author} {\bibfnamefont {W.}~\bibnamefont
  {Ku}}, \bibinfo {author} {\bibfnamefont {T.}~\bibnamefont {Berlijn}},\ and\
  \bibinfo {author} {\bibfnamefont {C.-C.}\ \bibnamefont {Lee}},\ }\bibfield
  {title} {\bibinfo {title} {Unfolding first-principles band structures},\
  }\href {https://doi.org/10.1103/PhysRevLett.104.216401} {\bibfield  {journal}
  {\bibinfo  {journal} {Phys. Rev. Lett.}\ }\textbf {\bibinfo {volume} {104}},\
  \bibinfo {pages} {216401} (\bibinfo {year} {2010})}\BibitemShut {NoStop}%
\bibitem [{\citenamefont {Xian}\ \emph {et~al.}(2022)\citenamefont {Xian},
  \citenamefont {Wang}, \citenamefont {Nie}, \citenamefont {Li}, \citenamefont
  {Han}, \citenamefont {Lin}, \citenamefont {Zhang}, \citenamefont {Liu},
  \citenamefont {Zhang}, \citenamefont {Miao} \emph {et~al.}}]{41}%
  \BibitemOpen
  \bibfield  {author} {\bibinfo {author} {\bibfnamefont {J.-J.}\ \bibnamefont
  {Xian}}, \bibinfo {author} {\bibfnamefont {C.}~\bibnamefont {Wang}}, \bibinfo
  {author} {\bibfnamefont {J.-H.}\ \bibnamefont {Nie}}, \bibinfo {author}
  {\bibfnamefont {R.}~\bibnamefont {Li}}, \bibinfo {author} {\bibfnamefont
  {M.}~\bibnamefont {Han}}, \bibinfo {author} {\bibfnamefont {J.}~\bibnamefont
  {Lin}}, \bibinfo {author} {\bibfnamefont {W.-H.}\ \bibnamefont {Zhang}},
  \bibinfo {author} {\bibfnamefont {Z.-Y.}\ \bibnamefont {Liu}}, \bibinfo
  {author} {\bibfnamefont {Z.-M.}\ \bibnamefont {Zhang}}, \bibinfo {author}
  {\bibfnamefont {M.-P.}\ \bibnamefont {Miao}}, \emph {et~al.},\ }\bibfield
  {title} {\bibinfo {title} {{Spin mapping of intralayer antiferromagnetism and
  field-induced spin reorientation in monolayer CrTe$_2$}},\ }\href
  {https://doi.org/10.1038/s41467-021-27834-z} {\bibfield  {journal} {\bibinfo
  {journal} {Nature communications}\ }\textbf {\bibinfo {volume} {13}},\
  \bibinfo {pages} {1} (\bibinfo {year} {2022})}\BibitemShut {NoStop}%
\end{thebibliography}
\end{document}